# Laminar Families and Metric Embeddings: Non-bipartite Maximum Matching Problem in the Semi-Streaming Model

Kook Jin Ahn    Sudipto Guha[*]

**Abstract**

In this paper, we study the non-bipartite maximum matching problem in the semi-streaming model. The maximum matching problem in the semi-streaming model has received a significant amount of attention lately. While the problem has been somewhat well solved for bipartite graphs, the known algorithms for non-bipartite graphs use $2^{\frac{1}{\epsilon}}$ passes or $n^{\frac{1}{\epsilon}}$ time to compute a $(1-\epsilon)$ approximation. In this paper we provide the first FPTAS (polynomial in $n, \frac{1}{\epsilon}$) for the problem which is efficient in both the running time and the number of passes. We also show that we can estimate the size of the matching in $O(\frac{1}{\epsilon})$ passes using slightly superlinear space.

To achieve both results, we use the structural properties of the matching polytope such as the laminarity of the tight sets and total dual integrality. The algorithms are iterative, and are based on the fractional packing and covering framework. However the formulations herein require exponentially many variables or constraints. We use laminarity, metric embeddings and graph sparsification to reduce the space required by the algorithms in between and across the iterations. This is the first use of these ideas in the semi-streaming model to solve a combinatorial optimization problem.

[*]Department of Computer Information Sciences, University of Pennsylvania, PA 19104. Email {kookjin,sudipto}@cis.upenn.edu. Research supported in part by an NSF Award CCF-0644119 and IIS-0713267 and a gift from Google.

# 1 Introduction

In this paper we consider the following question: Suppose that we are given a weighted (non-bipartite) graph $G = (V, E, w)$ with $|V| = n$ and $|E| = m$. Can we design a FPTAS for computing a $(1 + \epsilon)$ approximate maximum matching using only $\tilde{O}(n)$ additional space (not counting the input list of edges) such that we make a few passes over the list of edges? This question arises naturally in the semi-streaming model of computation. In this model we are given near linear (in $n$) space and we are allowed to make a small number of passes over a graph which is specified as an arbitrary (and possibly adversarial) list of edges. This is one of several natural models for processing massive graphs and the maximum matching problem has received a significant amount of attention [9, 16, 7, 23, 8, 17, 2]. For bipartite graphs the problem is somewhat well understood, based on a sequence of results [9, 16, 7, 2]. The state of the art is a $O(\frac{1}{\epsilon^2} \log \frac{1}{\epsilon})$ pass algorithm to compute a $(1 - \epsilon)$ approximation to the weighted maximum matching problem using $\tilde{O}(n \cdot poly(\frac{1}{\epsilon}))$ space and $\tilde{O}(m \cdot poly(\frac{1}{\epsilon}))$ time. However the status for non-bipartite graphs has been significantly more complicated. The results on bipartite graphs do not extend beyond a $\frac{2}{3}$ approximation for the case of non-bipartite graphs. In the case of non-bipartite graphs, we need to satisfy a significantly larger set of constraints, which correspond to odd subsets of vertices. To achieve a $(1 - \epsilon)$ approximation, the current algorithms either use $2^{\frac{1}{\epsilon}}$ passes [16] or $n^{\frac{1}{\epsilon}}$ time [2]; both of these results try to identify and enumerate over subsets (or paths) of size $O(\frac{1}{\epsilon})$. The open question, therefore, is about designing an $(1 - \epsilon)$ approximation algorithm which is efficient (in time polynomial in both $n$ and $\frac{1}{\epsilon}$, as well as a small number of passes). The central difficulty in designing such an algorithm is that any such algorithm would have to handle exponentially many subsets. In this paper we achieve such an algorithm using several novel ideas. We believe that the techniques developed in this paper will be useful in other scenarios where we wish to manage exponentially many constraints in a space efficient manner.

The difference between bipartite and non-bipartite graphs has been one of the cornerstone results in combinatorial optimization and a very rich literature that describes the non-bipartite matching polytope exists (see Schrijver [21]). The central idea of this paper is that the celebrated structural properties, such as the laminarity of the sets and the total dual integrality of the matching polytope (Cunningham and Marsh, [21]) can be used to design space and pass efficient algorithms for maximum matching. In particular we show that a non-adaptive perturbation to the constraints preserves the laminarity of odd sets (we only require this property for small size sets) in the matching polytope. The known proof of the Cunningham and Marsh theorem requires an adaptive two-stage optimization, see [21]. In that proof, in the first stage an optimum matching solution is chosen, and in the second stage a quadratic function is minimized (while retaining optimality). Solving the first step of their process in a space efficient fashion is the main challenge in our context; and thus the ideas in that proof do not apply directly. Intuitively, however, their proof suggests that we try to juggle two different objectives. We achieve this with the idea of a carefully designed perturbation, and a construction of outer LP and an inner LP. Both of these are solved in a space efficient manner. We believe this construction is of interest to the matching problem, independent of the semi-streaming model.

This paper also highlights the interesting conundrum between solving the primal problem of constructing a matching and the dual problem of covering edges by odd sets. The latter provides an estimate of the former and yet fails to give a construction of the matching. It appears that the latter can be solved with significantly better parameters (in the number of passes), albeit using randomization. In fact the number of passes used for the estimation of the size of the matching in non-bipartite case improves upon the number of passes required to construct a matching for a bipartite graph. Such a discrepancy between the packing and covering formulations was also observed by Onak and Rubinfeld [17] in a somewhat different setting. In the context of this paper, we show that the covering problem (which estimates the size of the matching) can be solved in $p/\epsilon$ passes (which is independent of $n$) for any $p \geq 1$, using slightly superlinear ($n^{1+O(\frac{1}{p})}$) space in the semi-streaming model using both graph sparsification and the Johnson-Lindenstrauss lemma. Thus the paper also demonstrates the use of these ideas in a combinatorial optimization problem in the semi-streaming model. The small number of passes can be achieved because the dual formulation of matching has a very small number of constraints (in comparison to the space allowed) but exponentially many variables, in contrast to the primal formulation which uses exponentially many constraints (and $m$ variables). However, the apparent complexity this estimation algorithm increases in comparison to the algorithm which constructs the matching, since in the dual we wish to solve a subproblem with exponentially many variables in each single pass and require stronger certificates of progress in between different passes. We require the certificate to be sufficiently simple, such that we can "compress" multiple iterations of the framework using a single pass of the stream. Thus the matching problem illustrates a rich tradeoff of passes, running time and space, as a function of the structure of the problem constraints.

**Our Results:** In this paper we provide two main results.

- First, we show that we can construct a maximum weighted matching in $O(\frac{1}{\epsilon^7} \log \frac{1}{\epsilon} \log n)$ passes using a deterministic algorithm that uses $\tilde{O}((m+n^3) \cdot poly(\frac{1}{\epsilon}))$ time and $\tilde{O}(n \cdot poly(\frac{1}{\epsilon}))$ space (Theorem 7). Moreover, we show that after suitable transformations, and using a notion of "effective weights", the problem reduces to an adaptive sequence of weighted *bipartite* matching formulations. These bipartite matching instances can then be solved in small space and a small number of passes using previous results, such as in [2]. We note that the reduction of the non-bipartite matching to a (relaxation of) bipartite matching is likely of independent interest.

- Second, we consider the problem of estimating the weight of the maximum matching (which uses the dual linear program) and show that we can estimate the weight in $p/\epsilon$ passes (which is independent of n) using a randomized algorithm that uses $\tilde{O}((m+n^4) \cdot poly(\frac{1}{\epsilon}))$ time and $n^{1+O(\frac{1}{p})}$ space (Theorems 23 and 32). Note that the number of passes is better than all previous results for bipartite graphs. However, the comparison is somewhat inexact because those previous results on bipartite graphs construct an explicit matching as well.

**Our Techniques:** Both the results use (and slightly modify) the framework of approximately solving packing and covering linear problems of Plotkin, Shmoys, and Tardos [19]. There has been a long sequence of results for similar approaches [22, 11, 12, 10, 3], but the approach of [19] explicitly allows side constraints specified as a polytope. We rely strongly on the ability to specify these side constraints and their duals. In fact one of the major contributions of this paper is to indicate which constraints should be added as side constraints.

In the first result we alter the constraints of the matching polytope by adding small perturbations to each odd set constraint. We then use a strengthening of the framework of [19], by adding a stronger exponential penalty than is mandated in [19]. As a consequence we are able to show that the (most) violated constraints form a *laminar family*. This allows us to reduce the exponential number of constraints (corresponding to the odd sets) to a linear number of relevant constraints. However finding the subsets still requires a non-trivial algorithm based on the minimum odd cut – and moreover the perturbation and penalties have to be carefully balanced to allow the reduction to the minimum odd cut problem. Finally, both the perturbation and the penalty affect the rate of progress (number of passes) of the algorithm. Designing this perturbation function is a contribution of this paper and is likely to be useful elsewhere.

In the second result we consider the dual formulation with $m$ constraints and exponential number of variables. We add a side constraint that is only true for optimum solutions which satisfy *laminarity* – the existence of such solutions are guaranteed by the Cunningham–Marsh Theorem. The covering framework iteratively provides weights to the edges (which correspond to the dual of the dual) and solves a subsidiary problem in each iteration. The space issue arises in both contexts. To provide the weights, we embed the weights into L2-distances between $n$ vectors, and the vectors can be compressed in $\tilde{O}(n)$ space using the Johnson-Lindenstrauss lemma. The subsidiary problem now reduces to a minimum odd cut problem as a consequence of the added side constraint. However, the input to the subsidiary problem is still an $O(m)$ size (streaming) vector of weights. To solve the problem in small space, we use streaming graph sparsification [1] to reduce the input to $\tilde{O}(n\epsilon^{-2})$. This is akin to reducing the number of relevant constraints. This sparsified problem is solved using the multiplicative weights update framework described in Arora, Hazan and Kale [3]. While this approach appears to be convoluted, a benefit of the approach is that the number of steps of the multiplicative weight update algorithm can be compressed in a single sparsification step – and random sampling can be used to compress the overall number of passes to a fairly low number.

**Roadmap:** We present a brief overview of the fractional packing and covering framework described in [19] in Section 2. We also present a brief overview of the multiplicative weight update method of [3]. We then present the result on constructing the weighted maximum matching in non-bipartite graphs in Section 3. The cardinality estimation algorithm using constant number of passes is presented in Section 4. Finally, in Section 5 we show how to extend the estimation algorithm to weighted maximum matching.

## 2 Preliminaries

In this section, we explain two frameworks for solving linear programs. In Sections 2.2 and 2.3, we explain the frameworks for fractional packing and covering problems, respectively, as presented by Plotkin, Shmoys, and Tardos [20]. In Section 2.4, we explain the multiplicative weight update method surveyed by Arora, Hazan, and Kale [3]. These approaches share the similar high level ideas but have differences in details, especially in handling side constraints.

### 2.1 The Matching Polytope

**Definition 1.** *Given a graph $G = (V, E)$ and a matching $M$, let $\mathbf{y}^M$ be an indicator vector of edges in $M$, i.e., $y_{ij}$ is 1 if $(i, j) \in M$ and 0 otherwise. Then, the **matching polytope** of $G$ is a convex hull of $\mathbf{y}^M$ for all matchings in $G$.*

**Theorem 1.** *[21] For any graph $G = (V, E)$, the polytope defined by the following constraints is the matching polytope:*

$$\begin{array}{ll} \sum_{j:(i,j)\in E} y_{ij} \leq 1 & \text{for all } i \in V \qquad \text{(vertex constraints)} \\ \sum_{i,j \in U} y_{ij} \leq \left\lfloor \frac{|U|}{2} \right\rfloor & \text{for all } U \subseteq V \qquad \text{(set constraints)} \end{array}$$

Observe that the feasibility of the constraint for a set $U$ where $|U|$ is even follows from the feasibility of the vertex constraints, since each $y_{ij}$ is summed up twice for $i, j \in U$. Therefore it suffices to focus on *odd* (size) sets $U$. Observe that the same observation holds for approximate feasibility, where the right hand sides of the equations corresponding to the constraints are multiplied by $(1 - \epsilon)$. From the definition of the matching polytope and Theorem 1, we can construct a linear program for the maximum matching problem with integrality gap one. LP1 and LP2 are the primal and dual linear programs for the maximum weighted matching problem. The maximum cardinality matching problem is the case when $w_{ij} = 1$ for all $(i,j) \in E$ and have a similar linear program formulation (See LP7 and LP8).

$$
\begin{array}{ll}
\max & \sum_{i,j} w_{ij} y_{ij} \\
\text{s.t} & \sum_j y_{ij} \leq 1 \quad \forall i \\
& \sum_{i,j \in U} y_{ij} \leq \left\lfloor \frac{|U|}{2} \right\rfloor \quad \forall U \\
& y_{ij} \geq 0
\end{array}
\quad \text{(LP1)}
\qquad
\begin{array}{ll}
\min & \sum_i x_i + \sum_U z_U \\
\text{s.t} & x_i + x_j + \sum_{i,j \in U} z_U \geq w_{ij} \quad \forall (i,j) \in E \\
& x_i, z_U \geq 0
\end{array}
\quad \text{(LP2)}
$$

The matching polytope is a well-studied object. One of the celebrated results on the matching polytope (with relation to the maximum matching problem) is the Cunningham-Marsh theorem.

**Theorem 2** (The Cunningham-Marsh Theorem [5, 21]). *If all edge weights are integers, there exists an optimal solution for the maximum weighted matching linear program(LP1) such that all $x_i$ and $z_U$ are integers and $\{U | z_U > 0\}$ is laminar.*

Note that the theorem does **not** claim that $x_i, z_U$ are 0/1 – which is only true when $w_{ij} = 1$. This issue is important in solving the LP for the weighted matching case. The most accessible proof of laminarity that uses an adaptive two-stage optimization (See [21], volume A, pages 440–442). The proof finds an optimal solution that maximizes $\sum_U z_U |U|^2$ (among optimal solutions) and shows that such an optimal solution satisfies the laminarity. The integrality of the optimal solution follows from the laminarity.

## 2.2 A Framework for Fractional Packing

Plotkin, Shmoys, and Tardos [20] presented a framework for solving fractional packing and fractional covering problems. A fractional packing problem can be written as follows:

$$
\begin{array}{ll}
\min & \lambda \\
\text{s.t.} & \mathbf{A}\mathbf{x} \leq \lambda \mathbf{b} \\
& \mathbf{x} \in \mathcal{P}
\end{array}
$$

where $\mathcal{P}$ is a convex polytope. The algorithm assumes that $\min_{x \in \mathcal{P}} \mathbf{c}^T \mathbf{x}$ can be computed efficiently for all $\mathbf{c}$.

**Definition 2.** *The* **width parameter** *of $\mathcal{P}$ is defined as $\rho = \max_i \max_{\mathbf{x} \in \mathcal{P}} \mathbf{A}_i \mathbf{x} / \mathbf{b}_i$.*

---

**Algorithm 1** Framework for the fractional covering [20]

---
1: Let $\lambda_0 = \max_i \mathbf{A}_i \mathbf{x} / \mathbf{b}_i$.
2: Let $\kappa$ be a parameter with $\kappa \geq 4\lambda_0^{-1} \delta^{-1} \ln(2m\delta^{-1})$ and $\sigma = \frac{\delta}{4\kappa\rho}$.
3: **while** $\max_i \mathbf{A}_i \mathbf{x} / \mathbf{b}_i \geq (1-\delta)\lambda_0$ and $\mathbf{y}$ do not satisfy (P2) **do**
4: $\quad \mathbf{y}_i \leftarrow \frac{1}{\mathbf{b}_i} \exp(\kappa \mathbf{A}_i \mathbf{x} / \mathbf{b}_i)$ for all $i$.
5: $\quad \tilde{\mathbf{x}} \leftarrow \operatorname{argmin}_{\mathbf{x} \in \mathcal{P}} \mathbf{y}^T \mathbf{A} \mathbf{x}$.
6: $\quad \mathbf{x} \leftarrow (1-\sigma)\mathbf{x} + \sigma \tilde{\mathbf{x}}$
7: **end while**

---

The framework (given in Algorithm 1) consists of multiple iterations of two-party game between the algorithm and an oracle. The algorithm maintains a primal candidate $\mathbf{x}$ and computes a dual candidate $\mathbf{y}$ (using exponential weights). Then, the oracle solves a subproblem given $\mathbf{y}$(line 5) and returns the solution to the algorithm. The algorithm updates its primal candidate and proceeds to the next iteration. In this process, the algorithm is fixed but the oracle varies depending on the problem. So we need to design a problem-specific oracle in order to use the framework.

**Definition 3.** *[20] Let $\lambda^{OPT}$ be the optimal solution of the fractional packing problem. Let $\lambda = \max_i \mathbf{A}_i\mathbf{x}/\mathbf{b}_i$ and $C_P(\mathbf{y}) = \min_{\mathbf{x}\in\mathcal{P}} \mathbf{y}^T\mathbf{A}\mathbf{x}$. $\mathbf{x}$ is $\delta$-optimal if $\mathbf{x}\in\mathcal{P}$ and $\lambda \leq (1+\delta)\lambda^{OPT}$.*

$$(P1) \quad (1-\delta)\lambda \mathbf{y}^T\mathbf{b} \leq \mathbf{y}^T\mathbf{A}\mathbf{x}$$
$$(P2) \quad \mathbf{y}^T\mathbf{A}\mathbf{x} - C_P(\mathbf{y}) \leq \delta(\mathbf{y}^T\mathbf{A}\mathbf{x} + \lambda \mathbf{y}^T\mathbf{b})$$

**Lemma 3.** *[20] If $\mathbf{x}$ and $\mathbf{y}$ satisfies (P1) and (P2) with $\delta \leq 1/6$, $\mathbf{x}$ is $6\delta$-optimal. Moreover, let $\Phi = \mathbf{y}^T\mathbf{b}$ be a potential function and let $\hat{\Phi}$ be the potential function after the update. Then, $\hat{\Phi} \leq (1-\kappa\sigma\delta\lambda)\Phi \leq (1-\Omega(\frac{\delta^2\lambda_0}{\rho}))\Phi$.*

For any $\mathbf{x}$, its corresponding $\mathbf{y}$ in Algorithm 1 satisfies (P1). Therefore, if (P2) is also satisfied, then we have a proof that $\mathbf{x}$ is near-optimal. If on the other hand, (P2) is not satisfied then the algorithm shows that the potential drops significantly. Initially, $\Phi \leq m\exp(\kappa\lambda_0)$ and the algorithm terminates if $\Phi \leq \exp((1-\delta)\kappa\lambda_0)$. Combined with the above lemma, we obtain the following:

**Theorem 4.** *Given the initial solution with objective value $\lambda_0$, we find a solution with objective value $(1-\delta)\lambda_0$ in $O(\frac{\kappa\rho}{\delta})$ iterations.*

Theorem 4 holds even if we find $\tilde{\mathbf{x}}$ approximately, i.e., find $\tilde{\mathbf{x}}$ such that $\mathbf{y}^T\mathbf{A}\tilde{\mathbf{x}} \leq (1+\delta/2)C_P(\mathbf{y}^t) + (\delta/2)\lambda\mathbf{y}^T\mathbf{b}$ [20]. It is also sufficient to compute $\mathbf{y}^T\mathbf{A}\mathbf{x}$ and $\lambda\mathbf{y}^T\mathbf{b}$ approximately (within a $1-\delta$ factor) based on the same ideas. Note that there is a slight difference between our parameters and the parameters in [20] which uses the smallest possible $\kappa$, namely $\kappa = 4\lambda_0^{-1}\delta^{-1}\ln(2m\delta^{-1})$. We will use a larger value for $\kappa$ in Section 3. This allows us to use structural properties of the polytope and subsequently allows us to approximate $\tilde{\mathbf{x}}$ easily.

## 2.3 A Framework for Fractional Covering Problems

The framework for fractional covering problems is almost identical to the framework for fractional packing problems except two differences. First, we use $\mathbf{y} = \frac{1}{b_i}\exp(-\kappa\mathbf{A}_i\mathbf{x}/\mathbf{b}_i)$. Note that the exponent is negative rather than positive. Second, (P1) and (P2) conditions are replaced by the following conditions.

**Definition 4.** *[20] Let $\lambda = \min_i \mathbf{A}_i\mathbf{x}/\mathbf{b}_i$ and $C_C(\mathbf{y}) = \min_{\mathbf{x}\in\mathcal{P}} \mathbf{y}^T\mathbf{A}\mathbf{x}$. $\mathbf{x}$ is $\delta$-optimal if $\mathbf{x}\in\mathcal{P}$ and $\lambda \geq (1-\delta)\lambda^*$.*

$$(C1) \quad (1+\delta)\lambda\mathbf{y}^T\mathbf{b} \geq \mathbf{y}^T\mathbf{A}\mathbf{x}$$
$$(C2) \quad C_C(\mathbf{y}) - \mathbf{y}^T\mathbf{A}\mathbf{x} \leq \delta(\mathbf{y}^T\mathbf{A}\mathbf{x} + \lambda\mathbf{y}^T\mathbf{b})$$

The subroutine to find $\tilde{\mathbf{x}}$ is a maximization problem given the cost $\mathbf{y}$, i.e., $\arg\max_{\mathbf{x}\in\mathcal{P}} \mathbf{y}^T\mathbf{A}\mathbf{x}$. The rest of the algorithm and proofs is almost identical to the fractional packing framework. We have the following theorem.

**Theorem 5.** *[20] If the initial solution is $6\epsilon$-optimal for $\epsilon \leq 1/12$, we find a $3\epsilon$-optimal solution in $O(\epsilon^{-2}\rho\log(m\epsilon^{-1}))$ iterations.*

## 2.4 The Multiplicative Weights Update Method

Similar to the frameworks for the fractional packing and the fractional covering problems, the multiplicative weights update method [3] consists of multiple iterations of two-party game between the algorithm and an oracle. A (minimization) LP and its dual program can be written in the following conventional forms where $\mathbf{A} \in \mathbb{R}^{m \times n}, \mathbf{b} \in \mathbb{R}^n, \mathbf{c} \in \mathbb{R}^m$:

$$\text{LP:} \begin{cases} \min & \mathbf{b}^T \mathbf{x} \\ \text{s.t} & \mathbf{A}^T \mathbf{x} \geq \mathbf{c}, \quad \mathbf{x} \geq \mathbf{0} \end{cases} \qquad \text{Dual LP:} \begin{cases} \max & \mathbf{c}^T \mathbf{y} \\ \text{s.t} & \mathbf{A} \mathbf{y} \leq \mathbf{b}, \quad \mathbf{y} \geq \mathbf{0} \end{cases}$$

The algorithm maintains weights $u_i^t$ for dual constraints $\mathbf{A}_i \mathbf{y} \geq b_i$ (these weights correspond to a candidate for a feasible primal solution) and the oracle returns a dual witness $\mathbf{y}^t$ or declare a failure. which reduces the duality gap (see the definition of admissibility below). If the oracle succeeds and returns a dual witness, it improves the candidate solution and if the oracle fails to return a dual witness, it proves that the primal solution is (near) optimal.

---

**Algorithm 2** The Multiplicative Weights Update Meta-Method [3]

1: $u_i^1 = 1$ for all $i \in [n]$.
2: **for** $t = 1$ **to** $T$ **do**
3:     Given $u_i^t$, the oracle returns a dual witness $\mathbf{y}^t$.
4:     Let $\mathrm{M}(i, \mathbf{y}^t) = \mathbf{A}_i \mathbf{y}^t - b_i$.
5:     Assert $-\ell \leq \mathrm{M}(i, \mathbf{y}^t) \leq \rho$.
6:     $u_i^{t+1} = \begin{cases} u_i^t (1 + \epsilon)^{\mathrm{M}(i, \mathbf{y}^t)/\rho} & \text{if } \mathrm{M}(i, \mathbf{y}^t) \geq 0 \\ u_i^t (1 - \epsilon)^{-\mathrm{M}(i, \mathbf{y}^t)/\rho} & \text{if } \mathrm{M}(i, \mathbf{y}^t) < 0 \end{cases}$
7: **end for**
8: Output $\frac{1}{T} \sum_t \mathbf{y}^t$ (along with any correction induced by failure of the oracle).

---

We have the following definition of dual witness and the corresponding theorem.

**Definition 5.** *We define $\mathrm{M}(i, \mathbf{y}^t) = \mathbf{A}_i \mathbf{y}^t - b_i$ to be a* **violation** *for Dual constraint $i$. The* **expected violation** $\mathrm{M}(\mathcal{D}^t, \mathbf{y}^t)$ *is the expected value of $\mathrm{M}(i, \mathbf{y}^t)$ when choosing $i$ with probability proportional to $u_i^t$, i.e., $\sum_i \frac{u_i^t}{\sum_j u_j^t} \mathrm{M}(i, \mathbf{y}^t)$. The dual witness is defined to be* **admissible** *if it satisfies*

$$\mathrm{M}(\mathcal{D}^t, \mathbf{y}^t) \leq \delta, \quad \mathbf{c}^T \mathbf{y}^t \geq \alpha, \quad \text{and} \quad \mathrm{M}(i, \mathbf{y}^t) \in [-\ell, \rho] \text{ with } \ell \leq \rho$$

*$\rho$ is defined as the width parameter of the oracle. The parameters $\epsilon$ and $T$ depend on $\rho$, $\ell$, and the desired error bound $\delta$.*

**Theorem 6.** *[3] Let $\delta > 0$ be an error parameter. Suppose that $\mathrm{M}(i, \mathbf{y}^t) \in [-\ell, \rho]$ with $\ell \leq \rho$. Then, after $T = \frac{2\rho \ln(n)}{\delta \epsilon}$ rounds and using $\epsilon = \min\{\frac{\delta}{4\ell}, \frac{1}{2}\}$, for any constraint $i$ we have:* $(1 - \epsilon) \sum_t \mathrm{M}(i, \mathbf{y}^t) \leq \delta T + \sum_t \mathrm{M}(\mathcal{D}^t, \mathbf{y}^t)$ *and thus $\mathrm{M}(i, \frac{1}{T} \sum_t \mathbf{y}^t) \leq 2\delta$ (since $\mathrm{M}(\mathcal{D}^t, \mathbf{y}^t) \leq \delta$ for all $t$).*

## 3 The Maximum Matching Problem

Our algorithm STREAMMATCH proceeds as follows.

1. We first find a 6-approximation using the algorithm of [9]. We can now guess the optimal solution up to $(1 - \delta)$ factor using $O(\frac{1}{\delta})$ guesses.

2. For each guess $\alpha$ of the optimum solution, we formulate the problem into testing the feasibility of a fractional packing problem. There is a well-known formulation for the maximum matching problem in general graphs whose integrality gap is 1 based on odd set constraints. Using subsets of size at most $\frac{1}{\delta}$, the same relaxation provides a $(1-\delta)$ approximation – however, the formulation is still difficult to solve efficiently (in time polynomial in $n, \frac{1}{\delta}$, and small number of passes) in the semi-streaming model. The following is the altered formulation where $f(\ell) = -\delta^2 \ell^2/4$. This choice of $f()$ requires $\kappa = 4\lambda_0^{-1} \delta^{-3} \ln(2m'\delta^{-1})$ where $m' = n^{\frac{1}{\delta}}$.

$$\min \quad \lambda$$
$$\mathcal{P} \begin{cases} \sum_{i,j \in U} y_{ij} \leq \lambda \left( \frac{|U|-1}{2} + f(|U|) \right) & \text{for all odd sets } U \text{ with } |U| \leq \frac{1}{\delta} \\ \sum_j y_{ij} \leq 1 & \text{for all } i \\ \sum_{i,j} w_{ij} y_{ij} \geq \alpha \end{cases} \quad \text{(LP3)}$$

The width parameter of the formulation is $O(1)$. The function $f()$ can be thought of as adding a small perturbation to each odd set. Observe that the formulation restricts the size of the sets (otherwise the perturbation would overwhelm the constraint). We discuss the details in Section 3.1.

3. We then consider applying the fractional packing framework (discussed in Section 2.2). The framework requires the computation of $\lambda$ which is defined over $n^{\frac{1}{\delta}}$ constraints. Moreover, the framework assigns $n^{\frac{1}{\delta}}$ weights $z_U$ (corresponding to the constraint $U$ for each odd set in the formulation) as well. We replace $z_U$ by $z'_U$ where $\{U|z'_U \neq 0\}$ is a laminar family. We then show that computing $z'_U$ (and $\lambda$) exactly on this laminar family, suffices to produce a $(1 - O(\delta))$ approximation oracle. In the process of computing $z'_U, \lambda$, we need to solve the following subproblem where $g(\ell) = -\delta^2 \ell^2/2$.

$$\min_U \sum_{i \in U} \left[ \lambda^* \left( 1 + \frac{2g(\ell)}{\ell} \right) - \sum_j y_{ij} \right] + \sum_{i \in U} \sum_{j \notin U} y_{ij}$$

The choice of $g()$ ensures that the subproblem can be reduced to a polynomial number of minimum odd cut problems. Moreover, we are performing a parametric search where we search for sets of size $\ell$ and the function $g$ satisfies (among other conditions) that $b\frac{g(\ell)}{\ell} + f(\ell) - g(\ell)$ is a convex function of $\ell$ minimized at $\ell = b$. We present two algorithms: a simple but inefficient algorithm in Section 3.2 and an efficient but complex algorithm in Section 3.3.

4. Finally, we consider solving the oracle problem required by the framework. Given $z'_U$, the oracle finds $\mathbf{y} \in \mathcal{P}$ such that $\sum_{ij} y_{ij} \sum_{i,j} z'_U$ is (approximately) minimized.

   (a) We now switch the constraint $\sum_{i,j} w_{ij} y_{ij} \leq \alpha$ and the objective function, based on guessing the minimum value $\beta$ of the objective function.
   (b) We use the Lagrangian multiplier technique and obtain the following LP (which is solved for several values of $\gamma$),

$$\max \quad \sum_{i,j \in E} w_{ij} y_{ij} + \gamma(\beta - \sum_{i,j} y_{ij} \sum_{i,j \in U} z'_U)$$
$$\sum_j y_{ij} \leq 1 \qquad \text{for all } i$$

Given $\gamma$, the LP objective function rewrites to $\gamma\beta + \sum_{i,j} \left( w_{ij} - \gamma \sum_{U: i,j \in U} z'_U \right) y_{ij}$. This is identical to the LP formulation of maximum weighted matching in bipartite graphs where we use an "effective weight" of $w_{ij} - \gamma \sum_{U: i,j \in U} z'_U$. Note that we are **not** saying that the graph is bipartite – just that this LP has a simple structure. These effective weights can be computed by a streaming algorithm and there is a semi-streaming algorithm for this LP [2].

We enumerate all possible values of $\gamma$ (we show that only a small number of values matter) and solve the LPs in parallel. Based on an appropriate choice of (a linear combination of) $\gamma$ we get $\beta = \sum_{i,j} y_{ij} \sum_{i,j \in U} z'_U$. We check if the objective function is larger than $(1 - O(\delta))\alpha$. Therefore we can solve the overall oracle approximately. This is explained in Section 3.4

As a consequence of the above we can show the following:

**Theorem 7.** *There is a $(1 - \epsilon)$-approximation algorithm that runs in $O(\frac{1}{\epsilon^7}(\log \frac{1}{\epsilon})(\log n))$ passes, $\tilde{O}(\frac{n}{\epsilon^8})$ space, and $\tilde{O}(\frac{1}{\epsilon^8}(m+\tau))$ time where $\tau$ is the time for finding a minimum odd cut over $\tilde{O}(\frac{n}{\epsilon^7})$ edges ($\tau = \tilde{O}(\frac{n^3}{\epsilon^7})$ suffices).*

**Roadmap:** The fractional packing framework is shown in Section 3.1. We show how to approximate $\lambda$ and $z_U$ is Sections 3.2 and 3.3. We describe the Lagrangian relaxation based solution in Section 3.4. Finally we summarize the discussion in the proof of Theorem 7 in Section 3.5.

## 3.1 Fractional Packing Formulation

We have the following LP for the maximum weighted matching in general graphs, see LP1 in Section 2.1.

$$
\begin{array}{ll}
\max & \sum_{i,j} w_{ij} y_{ij} \\
& \sum_j y_{ij} \leq 1 & \text{for all } i \\
& \sum_{i,j \in U} y_{ij} \leq \frac{|U|-1}{2} & \text{for all odd sets } U
\end{array}
$$

Let $f(\ell) = -\delta^2 \ell^2 / 4$. We alter LP1 as follows:

$$
\begin{array}{ll}
\max & \sum_{i,j} w_{ij} y_{ij} \\
& \sum_j y_{ij} \leq 1 & \text{for all } i \\
& \sum_{i,j \in U} y_{ij} \leq \frac{|U|-1}{2} + f(|U|) & \text{for all odd sets } U \text{ with } |U| \leq \frac{1}{\delta}
\end{array} \quad (\text{LP4})
$$

For the rest of this section, $U$ refers an odd set of size at most $\frac{1}{\delta}$ unless mentioned otherwise. We show that LP4 approximates LP1.

**Lemma 8.** *Let $OPT$ be the optimal solution for LP1. There exists a feasible solution for LP4 with objective value at least $(1-\delta)OPT$. Moreover if there is a feasible solution for LP4 with objective value $\alpha$, there is a feasible solution for LP1 with objective value $(1-\delta)\alpha$.*

*Proof.* For the first part of the lemma, let **y** be the optimal solution for LP1 and let $y'_{ij} = (1-\delta)y_{ij}$. The objective value of **y'** is $(1-\delta)OPT$. From the definition of $f$ and $|U| \leq \frac{1}{\delta}$, $f(|U|) \geq -\delta \frac{|U|-1}{2}$. Therefore $\sum_{i,j \in U} y'_{ij} = (1-\delta) \sum_{i,j \in U} y_{ij} \leq (1-\delta)\frac{|U|-1}{2} \leq \frac{|U|-1}{2} + f(|U|)$.

For the second part of the lemma, let **y** be a feasible solution for LP4 with objective value $\alpha$. Let $y'_{ij} = (1-\delta)y_{ij}$. Then, the objective value of **y'** is $(1-\delta)\alpha$. Since we scale down the solution, the constraints corresponding to nodes and odd sets of size less than $\frac{1}{\delta}$ are satisfied by **y'**. For larger odd sets $U$, we have $\sum_{i,j \in U} y'_{ij} \leq \frac{1}{2} \sum_{i \in U} \sum_j y'_{ij} = \frac{1-\delta}{2} \sum_{i \in U} \sum_j y_{ij} \leq \frac{1-\delta}{2}|U| \leq \frac{|U|-1}{2}$. □

Now we formulate the problem as a fractional packing problem, which is LP3. The corresponding oracle is:

$$
\begin{array}{ll}
\min & \sum_{i,j} y_{ij} \sum_{i,j \in U} z_U \\
\mathcal{P} & \left\{ \begin{array}{l} \sum_j y_{ij} \leq 1 \quad \text{for all } i \\ \sum_{i,j} w_{ij} y_{ij} \geq \alpha \end{array} \right.
\end{array} \quad (\text{LP5})
$$

## 3.2 A Simple Algorithm for Estimating $z_U$ and $\lambda$

The difficulty in solving LP5 arises from the fact that the number of variables $z_U$ is large. There are $n^{\Theta(\frac{1}{\delta})}$ odd sets of size at most $\frac{1}{\delta}$. So even computing all $z_U$ would take exponential time in $\frac{1}{\delta}$. Fortunately, we only need an approximation solution for LP5 and most of $z_U$ contribute little to the objective value. So we replace $z_U$ by $z'_U$ which are defined as follows:

**Definition 6.** *For odd sets $U$ with $1 < |U| \leq \frac{1}{\delta}$, let $Y_U = \sum_{i,j \in U} y_{ij}$, $b_U = \frac{|U|-1}{2} + f(|U|)$, $\lambda_U = Y_U/b_U$. Let $\lambda = \min_U \lambda_U$. Let*

$$z'_U = \begin{cases} z_U & \text{if } \lambda_U \geq \lambda - \delta^3 \\ 0 & \text{otherwise} \end{cases}$$

*and $L = \{U : z'_U \neq 0\}$.*

In the rest of this section, we present an algorithm to compute $\lambda$ and $z'_U$ exactly. Lemma 9 shows that we can approximate LP5 using $z'_U$. Lemma 10 means that either we have a (near) optimal solution or $L = \{U | z'_U > 0\}$ is a laminar family. If $L$ is a laminar family, all non-zero $z'_U$ values can be stored in $\tilde{O}(n)$ space, *across all the iterations*. Based on these two lemmas, we prove Theorem 11 which reduces to computing the minimum odd cut – however we need to parametrize the minimization to explicitly avoid large sets. For the sake of continuity of the discussion we first state the lemmas and the theorem we prove, and provide their proofs subsequently.

**Lemma 9.** *For any $\mathbf{y}' \in \mathcal{P}$, $\sum_{ij} y'_{ij} \sum_{i,j \in U} z'_U$ approximates $\sum_{ij} y'_{ij} \sum_{i,j \in U} z_U$ within $(\delta/4)\lambda \sum_U z_U b_U$ additive error. In addition, $\lambda \sum_U z'_U b_U$ approximates $\lambda \sum_U z_U b_U$ within $(1-\delta)$ factor.*

**Lemma 10.** *If $\lambda > 1 + 2\delta$, $L$ is a laminar family for a sufficiently small $\delta$.*

**Theorem 11.** *There is an algorithm that finds $\lambda$ and $L$ in polynomial time and near linear space.*

### Proof of Lemma 9

As mentioned in Section 3.1, the parameter $\kappa$ was set as $\kappa = 4\lambda_0^{-1}\delta^{-1}\ln(2m'\delta^{-1})$ in [20], where $m'$ is the number of constraints. Note that $m' = n^{\Theta(\frac{1}{\delta})}$ for the matching formulation we have.

We use $\kappa = 4\lambda_0^{-1}\delta^{-3}\ln(2m'\delta^{-1})$. Intuitively, we increase the cost of constraints with large violation so that only a smaller number of constraints matter. However, the larger value of $\kappa$ will increase the number of iterations.

*Proof of Lemma 9.* Let $\Phi = \lambda \sum_U z_U b_U$. Then, $\lambda \sum_U z_U b_U \geq \frac{\lambda \exp(\kappa \lambda)}{2}$. For $U$ with $\lambda_U \leq \lambda - 2\delta^3$ ($z'_U = 0$), we have $z_U \leq 2\exp(\kappa(\lambda_U - \lambda) + \kappa\lambda) \leq 4\exp(-\kappa\delta^3)\Phi \leq 4(n^{-\frac{1}{\delta}})^2\Phi$. Therefore:

$$\sum_{ij} y'_{ij} \sum_{i,j \in U} z_U - \sum_{ij} y'_{ij} \sum_{i,j \in U} z'_U = \sum_{ij} y'_{ij} \sum_{i,j \in U} (z_U - z'_U) \leq \sum_U (z_U - z'_U) \sum_{i,j \in U} y'_{ij} \leq \frac{1}{\delta} \sum_{U: z'_U = 0} z_U$$

$$\leq \frac{1}{\delta} n^{\frac{1}{\delta}} 4(n^{-\frac{1}{\delta}})^2 \Phi \leq \frac{\delta}{4}\Phi$$

For $\lambda \sum_U z'_U b_U$, we have a similar inequality; namely

$$\lambda \sum_U z_U b_U - \lambda \sum_U z'_U b_U \leq \lambda \sum_{U: z'_U = 0} z_U b_U \leq \frac{\delta}{4}\Phi$$

Therefore the lemma follows. □

**Proof of Lemma 10**

Now we can focus on $U$ with $\lambda_U \geq \lambda - \delta^3$ and show that such odd sets form a laminar family. The intuition behind the proof of the laminarity (and the choice of $f(\ell)$) is from a proof of the Cunningham-Marsh theorem (See [21], volume A, page 440–442). For the sake of the proof, we extend the definition of $Y_U$, $b_U$, and $\lambda_U$ (see Definition 6) to odd sets of size at most $\frac{2}{\delta}$.

*Proof of Lemma 10.* Suppose that $\lambda \geq 1 + 2\delta$ and $L$ is not a laminar family. Then, there are $A, B \in L$ such that $A \cap B \neq \emptyset, A, B$. There are two cases depending on $|A \cap B|$.

**Case I: $|A \cap B|$ is even.** Let $Q_A = \sum_{i \in A \cap B} \sum_{j \in A-B} y_{ij}$ and $Q_B = \sum_{i \in A \cap B} \sum_{j \in B-A} y_{ij}$. Without loss of generality, assume that $Q_A \leq Q_B$. Let $C = A - B$ and $D = A \cap B$. Then,

$$Y_C = Y_A - Q_A - Y_D \quad \text{and} \quad Y_D \leq \frac{1}{2}\left(\sum_{i \in D}\sum_j y_{ij} - Q_A - Q_B\right) \leq \frac{|D|}{2} - \frac{Q_A + Q_B}{2}$$

Since $b_A = \frac{|A|-1}{2} + f(|A|)$, $b_C = \frac{|C|-1}{2} + f(|C|)$, $|D| = |A| - |C|$, we have

$$Y_C \geq Y_A - \frac{|D|}{2} - \frac{Q_A}{2} + \frac{Q_B}{2} \geq Y_A - \frac{|D|}{2}$$

$$\begin{aligned}
b_A - b_C &= \frac{|A|-1}{2} + f(|A|) - \frac{|C|-1}{2} - f(|C|) = \frac{|A|-|C|}{2} + (f(|A|) - f(|C|)) \\
&= \frac{|D|}{2} + \frac{\delta^2}{4}(|A|+|C|)(|A|-|C|) \geq (1-\delta)\frac{|D|}{2}
\end{aligned}$$

For the last inequality, we used the fact that $|A| + |C| \leq \frac{2}{\delta}$.

$$\lambda_C b_C \geq \lambda_A b_A - (1-\delta)^{-1}(b_A - b_C) \geq \lambda_A b_C + (\lambda_A - (1-\delta)^{-1})(b_A - b_C) \geq \lambda_A b_C + \frac{\delta}{2}(b_A - b_C).$$

The last inequality is from the assumption that $\lambda_A \geq \lambda - \delta^3 \geq 1 + 2\delta - \delta^3$. Since $b_C \leq \frac{1}{\delta}$ and $b_A - b_C > 1 - \delta$, we get $\lambda_C > \lambda_A + \delta^3$. But this contradicts the assumption that $\lambda_A \geq \lambda - \delta^3$ since $\lambda \geq \lambda_C$.

**Case II: $|A \cap B|$ is odd.** Let $C = A \cap B$ and $D = A \cup B$. Let $t = (|A| + |B|)/2 = (|C| + |D|)/2$, $p = ||A| - t|$, and $q = ||C| - t|$. Then, $q \geq p + 2$ and we have

$$\begin{aligned}
b_C + b_D &= \frac{|C|-1}{2} + \frac{|D|-1}{2} + f(|C|) + f(|D|) = \frac{|C|-1}{2} + \frac{|D|-1}{2} - \frac{\delta^2}{4}(|C|^2 + |D|^2) \\
&= \frac{|C|-1}{2} + \frac{|D|-1}{2} - \frac{\delta^2}{4}(2t^2 + 2q^2) \leq \frac{|A|-1}{2} + \frac{|B|-1}{2} - \frac{\delta^2}{4}(2t^2 + 2p^2) - 2\delta^2 = b_A + b_B - 2\delta^2.
\end{aligned}$$

Since $b_A + b_B \leq \frac{2}{\delta}$, $b_C + b_D \leq (1 - \delta^3)(b_A + b_B)$. Wlog, we assume that $\lambda_A \leq \lambda_B$. Then,

$$\lambda_C b_C + \lambda_D b_D = Y_C + Y_D = Y_A + Y_B = \lambda_A b_A + \lambda_B b_B \geq \lambda_A(b_A + b_B)$$

If both $\lambda_C$ and $\lambda_D$ is less than $\lambda_A + \delta^3$,

$$\lambda_C b_C + \lambda_D b_D \leq (\lambda_A + \delta^3)(b_C + b_D) \leq \lambda_A(1 + \delta^3)(1 - \delta^3)(b_A + b_B) < \lambda_A(b_A + b_B).$$

So $\lambda_C$ or $\lambda_D$ is greater than $\lambda_A + \delta^3$ which means that either $\lambda > \lambda_A + \delta^3$ or $\lambda_D > \lambda_A + \delta^3$ with $|D| > \frac{1}{\delta}$. The former contradicts the assumption that $\lambda_A \geq \lambda - \delta^3$. In the latter case, we have $Y_D \leq |D|/2$ and

$$b_D = \frac{|D|-1}{2} - \frac{\delta^2}{4}|D|^2 = \frac{|D|}{2}\left(1 - \frac{1}{|D|} - \frac{\delta^2}{4}|D|\right) < \frac{|D|}{2}\left(1 - \delta - \frac{\delta^2}{4}\frac{2}{\delta}\right) = \frac{|D|}{2}\left(1 - \frac{3}{2}\delta\right)$$

Hence, $\lambda_D = Y_D/b_D \leq 1+2\delta$ and $\lambda_A < \lambda_D - \delta^3 < 1+2\delta - \delta^3$ which contradicts the assumption that $\lambda \geq 1+2\delta$ and $\lambda_A \geq \lambda - \delta^3$. From the above two cases, $L$ is a laminar family if $\lambda \geq 1+2\delta$. □

Since $L$ forms a laminar family, we can store $L$ in $O(n)$ space. In addition, since $|U|$ is bounded by $\frac{1}{\delta}$, the computation for $\sum_{i,j \in U} z_U$ given an edge $(i,j)$ can be performed in $O(\frac{1}{\delta})$ time.

## Proof of Theorem 11

Finally, we present algorithms to compute $\lambda$ and $z'_U$. We use an auxiliary function $F(\lambda^*, \ell, U)$ to reduce the problem into the minimum odd cut problem. $F(\lambda^*, \ell, U)$ is defined as follows.

**Definition 7.** *Let $g(\ell) = -\frac{\delta^2 \ell^2}{2}$ and $h(\ell) = \frac{\delta^2 \ell^2}{4}$. Let*

$$F(\lambda^*, \ell, U) = \sum_{i \in U} \left[ \lambda^* \left(1 - \frac{2g(\ell)}{\ell}\right) - \sum_j y_{ij} \right] + \sum_{i \in U} \sum_{j \notin U} y_{ij}.$$

**Algorithm 3** Find $\text{argmin}_U F(\lambda^*, \ell, U)$
---
1: If $|V|$ is even, add a node to $V$ without any adjacent edge so that $|V|$ is odd.
2: Construct a graph $G' = (V \cup \{s\}, E', w)$ such that $E'$ contains

  (i) $(i,j)$ with $w_{ij} = y_{ij}$ and

  (ii) $(i,s)$ with $w_{is} = \lambda^* \left(1 - \frac{2g(\ell)}{\ell}\right) - \sum_j y_{ij}$.

3: Find a minimum odd cut $(U, V \cup \{s\} - U)$.
4: **return** $U$

Algorithm 3 returns an odd set $U$ that minimizes $F(\lambda^*, \ell, U)$. In fact, the cut value of $(U, V \cup \{s\} - U)$ is exactly $F(\lambda^*, \ell, U)$. We repeatedly use Algorithm 3 for finding an odd set $U$ such that $\lambda_U \geq \lambda^*$ and $|U| = \ell$. $F(\lambda^*, \ell, U)$ satisfies the following three properties. Lemma 12 shows that if there is such an odd set, Algorithm 3 finds an odd set. Lemma 13 and Lemma 14 shows that we always find a set $U$ with $|U| = \ell$ (given $\lambda^*$ close to $\lambda \geq 1+2\delta$).

**Lemma 12.** *If there exists a set $U$ with $|U| = \ell$ and $\lambda_U \geq \lambda^*$, $F(\lambda^*, \ell, U) < \lambda^*(1 - 2h(\ell))$.*

*Proof.* From the definition of $\lambda_U$, we have $\sum_{i,j \in U} 2y_{ij} = \lambda_U(|U| - 1 + 2f(\ell))$. Since $\lambda^*$ is less than $\lambda_U$, we have $\sum_{i,j \in U} 2y_{ij} > \lambda^*(|U| - 1 + 2f(\ell))$. Applying $\sum_{i,j \in U} 2y_{ij} = \sum_{i \in U} \sum_j y_{ij} - \sum_{i \in U} \sum_{j \notin U} y_{ij}$ and $f(\ell) = g(\ell) + h(\ell)$, we obtain the desired result. □

**Lemma 13.** *If $|U| \neq \ell$ and $\lambda^* \geq \lambda_U - \delta^3$, $F(\lambda^*, \ell, U) > \lambda^*(1 - 2h(\ell))$.*

*Proof.* We have $F(\lambda^*, \ell, U) - \lambda^*(1 - 2h(\ell)) = \lambda^*(|U| - 1 + |U|\frac{2g(\ell)}{\ell} + 2h(\ell)) - \sum_{i,j \in U} 2y_{ij}$. Let $\hat{f}_U(\ell) = |U|\frac{g(\ell)}{\ell} + h(\ell)$. Then, $\hat{f}_U(\ell) = \frac{\delta^2}{2}(\ell^2 - 2|U|\ell)$ which is a convex function minimized at $\ell = |U|$. In addition, $\hat{f}_U(|U|) = f(|U|)$ and $\hat{f}_U(\ell) \geq f(|U|) + 2\delta^2$ for $\ell \neq |U|$. So

$$|U| - 1 + |U|\frac{2g(\ell)}{\ell} + 2h(\ell) \geq |U| - 1 + 2f(|U|) + 4\delta^2 \geq (1 + 4\delta^3)(|U| - 1 + 2f(|U|)) = (1 + \delta^3)2b_U$$

Also $\sum_{i,j \in U} 2y_{ij} = 2\lambda_U b_U$. From $\lambda^* \geq \lambda_U - \delta^3 \geq \lambda_U(1 - \delta^3)$, we obtain

$$F(\lambda^*, \ell, U) - \lambda^*(1 + 2h(\ell)) \geq (1 - \delta^3)(1 + 4\delta^3)2\lambda_U b_U - 2\lambda_U b_U = (3\delta^3 - 4\delta^6)2\lambda_U b_U > 0$$

which is the desired result. □

**Lemma 14.** *If $|U| > \frac{1}{\delta}$ and $\lambda^* \geq 1 + 2\delta - \delta^3$, $F(\lambda^*, \ell, U) > \lambda^*(1 - 2h(\ell))$.*

*Proof.* $|U|\frac{2g(\ell)}{\ell} + 2h(\ell)$ is a convex function that is minimized at $\ell = |U|$. Hence, it is minimized at $\ell = \frac{1}{\delta}$ if $\ell \leq \frac{1}{\delta}$ and $|U| > \frac{1}{\delta}$. So we have

$$|U| - 1 + |U|\frac{2g(\ell)}{\ell} + 2h(\ell) = |U| - 1 - |U|\ell\delta^2 + \frac{\delta^2\ell^2}{2} \geq |U| - 1 - \delta|U| + \frac{1}{2} \geq |U|(1 - \frac{3}{2}\delta).$$

On the other hand, $\sum_{i,j \in U} 2y_{ij} \leq |U|$. Therefore,

$$F(\lambda^*, \ell, U) - \lambda^*(1 - 2h(\ell)) = \lambda^*(|U| - 1 + |U|\frac{2g(\ell)}{\ell} + 2h(\ell)) - \sum_{i,j \in U} 2y_{ij} \geq \lambda^*|U|(1 - \frac{3}{2}\delta) - |U|$$

The RHS of the above equation is $\geq ((1 + 2\delta - \delta^3)(1 - \frac{3}{2}\delta) - 1)|U| > 0$. □

Using the properties of $F(\lambda^*, \ell, U)$, Algorithm 4 and Algorithm 5 computes $\lambda$ and $L$ by enumerating all possible values of $\ell$.

---

**Algorithm 4** Find $\lambda$ given **y**.
1: $\lambda^* \leftarrow 1 + 2\delta$.
2: **for** $\ell = 3$ to $\frac{1}{\delta}$ **do**
3:     Find $\text{argmin}_U F(\lambda^*, \ell, U)$ in $G$.
4:     **if** $\lambda_U > \lambda^*$ **then**
5:         $\lambda^* \leftarrow \lambda_U$
6:         Repeat 3
7:     **end if**
8: **end for**
9: **return** $\lambda^*$.

**Algorithm 5** Find $L$ given **y** and $\lambda$.
1: Let $\lambda^* = \lambda - \delta^3$.
2: **for** $\ell = 3$ to $\frac{1}{\delta}$ **do**
3:     $G' \leftarrow G$.
4:     Find $\text{argmin}_U F(\lambda^*, \ell, U)$ in $G'$.
5:     **if** $|U| = \ell$ and $\lambda_U \geq \lambda - \delta^3$ **then**
6:         $L \leftarrow L \cup \{U\}$.
7:         Delete all nodes in $U$ (and adjacent edges) from $G'$.
8:         Repeat 4.
9:     **end if**
10: **end for**
11: **return** $L$

---

*Proof of Theorem 11.* First, we prove that Algorithm 4 finds $\lambda$ with $O(\frac{1}{\delta})$ minimum odd cut problems. Let $U_\ell = \text{argmax}_{|U|=\ell} \lambda_U$. By Lemma 12, if $\lambda_{U_\ell} > \lambda^*$, we find at least one $U$ with $\lambda_U > \lambda^*$. Therefore, Algorithm 4 finds $\lambda$ eventually. In addition, we find $U$ with $|U| = \ell$ only if $\lambda_U = \lambda_{U_\ell}$. So for each odd set size $\ell$, we update $\lambda^*$ at most once. Therefore, we compute a minimum odd set at most $\frac{2}{\delta}$ times.

Algorithm 5 finds all $U$ with $\lambda_U \geq \lambda - \delta^3$ in polynomial time. If there is an odd set $U \in L$ with $|U| = \ell$, we find an odd set by Lemma 12 and the size of the set is $\ell$ by Lemma 13 and Lemma 14. So if $|U| \neq \ell$, no set of size $\ell$ in $L$ is left in $G$. Therefore, we find all sets in $L$ by enumerating all possible values of $\ell$. The number of times we need to solve the minimum odd set is $O(n + \frac{1}{\delta})$ which is the desired result. □

### 3.3 An Efficient Algorithm for Estimating $z_U$

The bottleneck on the running time to compute $\lambda, z'_U$ is Algorithm 5. Algorithm 5 repeatedly finds $\text{argmin}_U F(\lambda^*, \ell, U)$ which solves a minimum odd cut. In Algorithm 5, we find one odd set at a time which causes $O(n)$ executions of the minimum odd cut algorithm. Padberg and Rao's minimum odd cut algorithm constructs a Gomory-Hu tree [18]. Instead of throwing out the Gomory-Hu tree after finding one odd set, we can reuse the Gomory-Hu tree using the laminarity of $L$ and find multiple odd sets from the tree. Algorithm 6 is the improved version of Algorithm 5. We prove that the algorithm finds at least half of $U \in L$ with $|U| = \ell$ per iteration.

**Algorithm 6** Find $L$ given $\mathbf{y}$ and $\lambda$.
1: Let $\lambda^* = \lambda - \delta^3$.
2: $L \leftarrow \emptyset$.
3: **for** $\ell = 3$ to $\frac{1}{\delta}$ **do**
4: $\quad \tilde{G} = (\tilde{V}, \tilde{E}) \leftarrow G$.
5: $\quad$ **repeat**
6: $\qquad$ If $|\tilde{V}|$ is even, add a node to $\tilde{V}$ without any adjacent edge so that $|\tilde{V}|$ is odd.
7: $\qquad$ Construct a graph $G' = (\tilde{V} \cup \{s\}, E', w)$ such that $E'$ contains

$\qquad$ (i) $\quad (i,j)$ with $w_{ij} = y_{ij}$ and

$\qquad$ (ii) $\quad (i,s)$ with $w_{is} = \lambda^* \left(1 - \frac{2g(\ell)}{\ell}\right) - \sum_j y_{ij}$.

8: $\qquad$ Construct a Gomory-Hu tree of $G'$.
9: $\qquad$ **for each** edge $e$ in the Gomory-Hu tree with $w_e \leq \lambda^*(1 - 2h(\ell))$ **do**
10: $\qquad\quad$ Let $(U, \tilde{V} \cup \{s\} - U)$ be the cut obtained by erasing $e$ from the tree.
11: $\qquad\quad$ **if** $|U| = \ell$ **then**
12: $\qquad\qquad$ $L \leftarrow L \cup \{U\}$.
13: $\qquad\qquad$ Delete all nodes in $U$ (and adjacent edges) from $\tilde{G}$.
14: $\qquad\quad$ **end if**
15: $\qquad$ **end for**
16: $\quad$ **until** No $U$ is added to $L$
17: $\quad$ **return** L
18: **end for**

**Lemma 15.** *Let $L_\ell = \{U | U \in L, |U| = \ell\}$. In Algorithm 6, for each iteration of line 5–16, we find at least half of remaining sets in $L_\ell$.*

*Proof.* We use the proof of the minimum odd cut algorithm in [21]. For the details of the proof, see [21], volume A, page 499. Let $F$ be the edge set of the Gomory-Hu tree. For $f \in F$, let $W_f$ be one of the two components obtained by erasing $f$ from the Gomory-Hu tree. Let $\delta_F(U)$ be the edges in $L$ which are cut by $(U, \tilde{V} \cup \{s\} - U)$.

Suppose that $U \in L_\ell$. Then, there must be $f = (u,v) \in \delta_F(U)$ such that $W_f$ is an odd set(from [21]). $W_f$ is a minimum $u-v$ cut and hence, the cut value is less than or equal to $U$. By Lemma 13 and Lemma 14, only odd sets of size $\ell$ has cut value less than $U$. Therefore, $W_f$ or its complement is in $L_\ell$. Let this odd set be $U'$. If $U \neq U'$, Algorithm 6 does not find $U$ but does find $U'$. Furthermore, the only way to find $U'$ is by choosing $f$. So $U'$ does not eliminate other sets in $L_\ell$. Therefore, the algorithm covers at least half of remaining $U \in L_\ell$ per each iteration of line 5–16. □

By repeating $O(\log n)$ times, we find all sets in $L_\ell$. By enumerating over all possible $\ell$ values, we obtain the following theorem which is Theorem 11 with the improved running time of Algorithm 6.

**Theorem 16.** *There is an algorithm that finds $L$ in $\tilde{O}(\frac{\tau}{\delta})$ time where $\tau$ is the time to construct a Gomory-Hu tree for an $n$ node graph with $\tilde{O}(n \cdot poly(\frac{1}{\delta}))$ edges (which is $n$ times the number of passes).*

### 3.4 Solving the Oracle with the Lagrangian Dual

Given $z'_U$, the oracle (approximately) decides if there exists $\tilde{\mathbf{y}} \in \mathcal{P}$ such that $\sum_{ij} \tilde{y}_{ij} \sum_{i,j} z'_U$ is less than $(1 - \delta) \sum_{ij} y_{ij} \sum_{i,j} z'_U - \delta \lambda \sum_U z'_U b_U$.

Let $\beta = (1-\delta)\sum_{ij} y_{ij} \sum_{i,j} z'_U - \delta\lambda \sum_U z'_U b_U$. Then, we can reduce the oracle LP5 into the following linear program – note that we have switched the objective function and the constraint since we guessed the objective function (as discussed earlier in Section 3). Note that we are interested in the solution of the above only if the new objective function is at least $(1 - O(\delta))\alpha$.

$$\begin{aligned} \max \quad & \sum_{i,j} w_{ij} y_{ij} \\ & \sum_j y_{ij} \leq 1 \quad \text{for all } i \\ & \sum_{i,j} y_{ij} \sum_{i,j \in U} z'_U \leq \beta \end{aligned}$$

Taking the Lagrangian relaxation of the final constraint, we obtain the following linear program.

$$\min_\gamma \max_\mathbf{y} \quad \sum_{i,j} w_{ij} y_{ij} + \gamma(\beta - \sum_{i,j} y_{ij} \sum_{i,j \in U} z'_U) \qquad \text{(LP6)}$$
$$\sum_j y_{ij} \leq 1 \quad \text{for all } i$$

Observe that the objective function of LP6 rearranges to $\gamma\beta + \sum_{i,j} \left(w_{ij} - \gamma \sum_{U:i,j \in U} z'_U\right) y_{ij}$. Note that $\beta\gamma > 0$ and therefore (approximating) LP6 is identical to the linear program for the bipartite maximum weighted matching problem except $\gamma\beta$ in the objective value. We presented an approximation algorithm for the bipartite maximum weighted matching problem [2].

**Theorem 17.** *[2] There is a $(1-\delta)$-approximation algorithm for the bipartite maximum weighted matching problem that runs in $O(\delta^{-2} \log \delta^{-1})$ passes, $\tilde{O}(n)$ space, and $\tilde{O}(m)$ time.*

**Corollary 18.** *We can find a $(1-\delta)$ approximation for LP6 using $O(\delta^{-2} \log \delta^{-1})$ passes, $\tilde{O}(n)$ space, and $\tilde{O}(m)$ time; where on seeing an edge $(i,j)$ we compute its "effective" weight to be $w_{ij} - \gamma \sum_{U:i,j \in U} z'_U$ and use the maximum weighted bipartite matching algorithm. Note that we discard all edges with negative effective weight.*

We use an approach similar to the solution of the linear programming relaxation of the $k$-median problem, as in Jain and Vazirani [14]. We enumerate possible values of $\gamma$ and approximate LP6. By taking a linear combination of two of them, we find an approximation solution for LP6. Algorithm 7 is the approximation algorithm for LP6.

**Definition 8.** *Let OPT be the optimal solution of LP6. Let $q_{ij} = \sum_{i,j \in U} z'_U$.*

---

**Algorithm 7** Find $\tilde{\mathbf{y}}$

1: **for** $k = 0$ to $\lceil \frac{9}{\delta} \rceil + 18$ **in parallel do**
2:    Let $\gamma = \frac{\delta\alpha k}{\beta}$.
3:    Find a $(1-\delta)$-approximation given $\gamma$ for LP6 given $\gamma$.
4:    Let $\mathbf{y}^\gamma$ be the approximation solution and $LP(\gamma)$ be the objective value of $\mathbf{y}_\gamma$.
5: **end for**
6: **if** $\sum_{ij} y^0_{ij} q_{ij} \leq \beta$ **then**
7:    return $\mathbf{y}^0$.
8: **else**
9:    Find $\gamma$ and $\gamma' = \gamma + \frac{\delta\alpha}{\beta}$ such that $\sum_{ij} y^\gamma_{ij} q_{ij} > \beta$ and $\sum_{ij} y^{\gamma'}_{ij} q_{ij} \leq \beta$.
10:    Let $a = \frac{\beta - \sum_{ij} y^{\gamma'}_{ij}}{\sum_{ij} y^\gamma_{ij} - \sum_{ij} y^{\gamma'}_{ij}}$.
11:    **return** $a\mathbf{y}^\gamma + (1-a)\mathbf{y}^{\gamma'}$.
12: **end if**

---

**Lemma 19.** *If $\sum_j y_{ij} \leq 1$ for all $i$, $\sum_{ij} w_{ij} y_{ij} = 9\alpha$.*

*Proof.* Let $M$ be the size of the maximum matching. We initially find a 6-approximation for $M$. The size initial matching is the lowerbound of $\alpha$. On the other hand, the integrality gap is $3/2$ which means the $\sum_{ij} w_{ij} y_{ij} \leq \frac{3}{2} M$. Therefore $\sum_{ij} w_{ij} y_{ij} = O(\alpha)$. □

**Lemma 20.** *Suppose that $\gamma \geq \frac{(1+2\delta)(9\alpha)}{\beta}$ and $\mathbf{y}^\gamma$ is the corresponding fractional matching returned by the bipartite matching algorithm. Then, $\sum_{ij} y_{ij}^\gamma q_{ij} \leq \beta$.*

*Proof.* If $\gamma \geq \frac{(1+2\delta)(9\alpha)}{\beta}$, then we can obtain a solution with objective value at least $(1+2\delta)(9\alpha)$ by assigning 0 to all variables. If $\sum_{ij} y_{ij}^\gamma q_{ij} > \beta$, the objective value is less than $\sum_{ij} w_{ij} y_{ij}^\gamma \leq 9\alpha$ by Lemma 19. This contradicts the fact that $\mathbf{y}^\gamma$ is a $(1-\delta)$-approximation given $\gamma$. Therefore, $\sum_{ij} y_{ij}^\gamma q_{ij} \leq \beta$. □

**Lemma 21.** *Algorithm 7 returns $\mathbf{y}$ such that $\sum_{ij} y_{ij} \geq (1-7\delta)OPT$ and $\sum_{ij} y_{ij} q_{ij} \leq \beta$.*

*Proof.* It is obvious that $LP(\gamma) \geq (1-\delta)OPT$ because we find a $(1-\delta)$-approximation for each $\gamma$. If $\sum_{ij} y_{ij}^0 q_{ij} \leq \beta$, the algorithm returns $\mathbf{y}^0$ and $\sum_{ij} y_{ij}^0 \geq (1-\delta)OPT$. If $\sum_{ij} y_{ij}^0 q_{ij} > \beta$, there must exist $\gamma$ and $\gamma' = \gamma + \frac{\delta\alpha}{\beta}$ such that $\sum_{ij} y_{ij}^\gamma > \beta$ and $\sum_{ij} y_{ij}^{\gamma'} \leq \beta$ by Lemma 20. By the construction of $\mathbf{y}$, $\sum_{ij} y_{ij} q_{ij} = \beta$. Then,

$$
\begin{aligned}
(1-\delta)OPT &\leq aLP(\gamma) + (1-a)LP(\gamma) \\
&= a\left[\sum_{ij} y_{ij}^\gamma + \gamma(\beta - \sum_{ij} y_{ij}^\gamma q_{ij})\right] + (1-a)\left[\sum_{ij} y_{ij}^{\gamma'} + \gamma'(\beta - \sum_{ij} y_{ij}^{\gamma'} q_{ij})\right] \\
&= a\left[\sum_{ij} y_{ij}^\gamma + \gamma(\beta - \sum_{ij} y_{ij}^\gamma q_{ij})\right] + (1-a)\left[\sum_{ij} y_{ij}^{\gamma'} + \gamma(\beta - \sum_{ij} y_{ij}^{\gamma'} q_{ij}) + (\gamma'-\gamma)(\beta - \sum_{ij} y_{ij}^{\gamma'} q_{ij})\right] \\
&\leq \sum_{ij} y_{ij} + \gamma(\beta - \sum_{ij} y_{ij} q_{ij}) + \frac{\delta\alpha}{\beta}\beta = \sum_{ij} y_{ij} + \delta\alpha.
\end{aligned}
$$

Since $\alpha \leq 6OPT$, $\sum_{ij} w_{ij} y_{ij} \geq (1-7\delta)OPT$. □

Hence, if $\alpha$ is less than $(1-7\delta)$ fraction of the maximum matching, the oracle finds a $\tilde{\mathbf{y}} \in P$ such that improves the solution significantly. Combining all the components, we obtain a $(1 - O(\delta))$-approximation of the maximum weighted matching. Summarizing the entire discussion, we obtain the proof of Theorem 7.

### 3.5 Proof of Theorem 7

*Proof.* The width parameter of LP5 is $O(1)$ and the number of constraints is $O(n^{\frac{1}{\epsilon}})$. Combined with our choice of $\kappa$, the total number of iterations for the fractional packing framework is $O(\frac{1}{\epsilon^5} \log n)$. For each iteration, we compute $\lambda$ and $z_U'$ which takes $\tilde{O}(\frac{\tau}{\epsilon})$ time where $\tau$ is the time for constructing a Gomory-Hu tree. (Using the preflow-push algorithm [13], it takes $\tilde{O}(n^2 m')$ time for constructing a Gomory-Hu tree where $m'$ is the number of edges. In our case, $m' = \tilde{O}(\frac{n}{\epsilon^7})$ and hence, $\tau = \tilde{O}(\frac{n^3}{\epsilon^7})$ suffices.) For LP6, it takes $O(\frac{1}{\epsilon^2} \log \frac{1}{\epsilon})$ passes and $\tilde{O}(\frac{m}{\epsilon^2})$ time. Since we execute the algorithm for $O(\frac{1}{\epsilon})$ guesses of the optimal solution in parallel, we obtain the desired number of passes, space, and time. Note that the space is dominated by the total number of edges chosen by the oracle. Each oracle execution returns at most $\tilde{O}(\frac{n}{\epsilon^2})$ edges and therefore, the required space is $\tilde{O}(\frac{n}{\epsilon^7})$. □

# 4 Estimating the Size of the Maximum Cardinality Matching

In this section, we present an algorithm that approximates the size of the maximum cardinality matching problem (MCM) in $\tilde{O}(\frac{p}{\epsilon})$ passes, $\tilde{O}(n^{1+1/p})$ space and polynomial time. The algorithm solves the dual linear program of MCM in general graphs and only approximates the size of the optimal matching. LP7 and LP8 are the primal and dual program for MCM in general graphs. Note that LP7 is equivalent to LP1 with $w_{ij} = 1$ for all $(i, j)$.

$$
\begin{array}{ll}
\max & \sum_{(i,j) \in E} y_{ij} \\
\text{s.t} & \sum_{(i,j) \in E} y_{ij} \leq 1 \quad \forall i \\
& \sum_{(i,j) \in U} y_{ij} \leq \left\lfloor \frac{|U|}{2} \right\rfloor \quad \forall U \\
& y_{ij} \geq 0
\end{array}
\quad \text{(LP7)}
\qquad
\begin{array}{ll}
\min & \sum_i x_i + \sum_U z_U \\
\text{s.t} & x_i + x_j + \sum_{i,j \in U} z_U \geq 1 \quad \forall (i,j) \in E \\
& x_i, z_U \geq 0
\end{array}
\quad \text{(LP8)}
$$

We first use the following simple algorithm SIMSTREAMMATCHEST and subsequently show how to improve the number of passes taken by the algorithm.

**Algorithm:** SIMSTREAMMATCHEST

1. As in Section 3.1, we bound the size of odd sets to $\frac{1}{\delta}$. The resulting linear program $(1 - \delta)$-approximates LP7. and has at most $n^{\frac{1}{\delta}}$ constraints.

2. We formulate the dual linear program as a fractional covering problem with a small width parameter.

$$
\begin{array}{ll}
\max & \lambda \\
\text{s.t} & x_i + x_j + \sum_{i,j \in U} z_U \geq \lambda \quad \forall (i,j) \in E \\
\mathcal{P} & \left\{ \begin{array}{l} \sum_i x_i + \sum_U \lfloor \frac{|U|}{2} \rfloor z_U \leq \alpha \\ 2x_i + \sum_{i \in U} z_U \leq 3 \quad \forall i \end{array} \right.
\end{array}
\quad \text{(LP9)}
$$

In particular, we use the Cunningham-Marsh theorem to show that we can add $2x_i + \sum_{i \in U} z_U \leq 3$ to LP8 without changing the optimal objective value. Note that $\sum_{i,j \in U} z_U \leq \sum_{i \in U} z_U$ and hence the width parameter is 4.

3. The oracle for LP9 solves the following problem given $y_{ij}$. LP10 is the oracle and LP11 is its dual LP.

$$
\begin{array}{ll}
\max & \sum_i x_i \sum_j y_{ij} + \sum_U z_U \sum_{i,j \in U} y_{ij} \\
\text{s.t} & \frac{1}{\alpha}[\sum_i x_i + \sum_U \lfloor \frac{|U|}{2} \rfloor z_U] \leq 1 \\
& 2x_i + \sum_{i \in U} z_U \leq 3 \quad \forall i
\end{array}
\quad \text{(LP10)}
\qquad
\begin{array}{ll}
\min & \sum_i 3\zeta_i + \zeta_\alpha \\
\text{s.t} & 2\zeta_i + \frac{1}{\alpha}\zeta_\alpha \geq \sum_j y_{ij} \quad \forall i \\
& \sum_{i \in U} \zeta_i + \lfloor \frac{|U|}{2} \rfloor \frac{1}{\alpha} \zeta_\alpha \geq \sum_{i,j \in U} y_{ij} \quad \forall U
\end{array}
\quad \text{(LP11)}
$$

If we have an oracle that approximates the above subproblem(LP10) in near linear space and polynomial time, we can construct an algorithm for LP8 using the fractional covering framework of [20]. However achieving such an oracle is nontrivial and we describe how to design an approximate oracle in the next few steps.

4. To solve the oracle problem given in LP10, we observe that the bottleneck arises in two steps. First, the weights $y_{ij}$ cannot be stored (since they are $m$ in number) and therefore the objective function of LP10 itself is an issue. Second, the LP has to be solved in a space efficient manner. Third, the solution has to be stored in a space efficient manner such that the $\sum_{i,j\in U} z_U$ can be computed when we generate $y_{ij}$ (for the next iteration). We address these issues as follows:

(a) First, we sparsify $y_{ij}$ so that $\tilde{O}(n)$ edges has non-zero values while approximating the optimal solution. This can be done with a semi-streaming sparsification algorithm [1]. We store the sparsification and therefore have random access to it.

(b) We then solve the LP10 with the multiplicative weights updated method [3]. In each iteration, we find a violated constraint in LP11 and construct a dual witness using the constraint. We can also use the ellipsoid method to solve LP11 using violated constraints (however we need to be careful about the next step).

(c) However, we need a near linear size sketch of a solution for LP10 and use the Johnson-Lindenstrauss Lemma to sketch $\sum_{i,j\in U} z_U$. Note that we keep updating this sketch as we keep on generating $U, z_U$ since storing these sets till the end of the oracle need not be space efficient.

Combining the above components, we obtain the following theorem.

**Theorem 22.** *The semi-streaming algorithm* SIMSTREAMMATCHEST *approximates the size of the maximum cardinality matching to a $(1-\epsilon)$ factor in $O(\frac{\log n}{\epsilon^2})$ passes, $\tilde{O}(\frac{n}{\epsilon^4})$ space, and $\tilde{O}(\frac{n\tau}{\epsilon^6})$ time where $\tau$ is the time to compute minimum odd cut over a graph with $\tilde{O}(\frac{n}{\delta^2})$ edges ($\tau = \tilde{O}(\frac{n^3}{\delta^2})$ suffices).*

**Space Pass Tradeoffs and Constant Number of Passes:** The algorithm SIMSTREAMMATCHEST uses a number of passes which depends on $n$. In Section 4.6 we describe a method STREAMMATCHEST to process multiple iterations (of the overall framework) in one pass.

**Theorem 23.** *The semi-streaming algorithm* STREAMMATCHEST *approximates the size of the maximum cardinality matching to a $(1-\epsilon)$ factor in $O(\frac{p}{\epsilon})$ passes, $\tilde{O}(\frac{n^{1+1/p}}{\epsilon^4})$ space, and $\tilde{O}(\frac{n\tau}{\epsilon^6})$ time where $\tau$ is the time to compute minimum odd cut over a graph with $\tilde{O}(\frac{n}{\epsilon^2})$ edges ($\tau = \tilde{O}(\frac{n^3}{\epsilon^2})$ suffices).*

**Roadmap:** In Section 4.1 we show that the addition of the constraint does not alter the problem. In Section 4.2, we prove that an approximation solution given the sparsification of $y_{ij}$ is an approximation solution of LP10 given $y_{ij}$. In Section 4.3, we describe a polynomial time algorithm that finds a violated constraint. In Section 4.4, we present an algorithm that sketches $z_U$; the algorithm produces a sketch that estimates $\sum_{i,j\in U} z_U$ for all $i,j$ pairs. Theorem 22 is proved in Section 4.5. In Section 4.6, we describe the pass space tradeoffs and prove Theorem 23.

## 4.1 The Fractional Covering Formulation

We use the Cunningham-Marsh Theorem (Section 2.1, Theorem 2 to show that we can add $2x_i + \sum_{i\in U} z_U \leq 3$ to LP8 without changing the optimal objective value. This is based on the following straightforward corollary (for $w_{ij} = 1$ only).

**Corollary 24.** *There exists an optimal solution for LP8 such that all $x_i$ and $z_U$ are 0 or 1 and $U$ with $z_U = 1$ are disjoint sets. As a consequence we can add $2x_i + \sum_{i\in U} z_U \leq 3$ to LP8 without changing the optimal objective value.*

*Proof.* If $x_i$ or $z_U$ is 1, it satisfies all the constraints that constrains the variable. So it is obvious that $x_i$ and $z_u$ are 0 or 1 in an optimal integral solution. If $z_U = 1$, it satisfies all the constraints that correspond to $(i, j) \in U'$ for $U' \subset U$. Hence, $z_{U'} = 0$ for all $U' \subsetneq U$ in the integral optimal solution. Therefore, if $\{U | z_U = 1\}$ is laminar in an optimal solution, such sets are disjoint.

The feasibility of the additional constraint follows immediately. □

## 4.2 Sparsification of $y_{ij}$, the input to the Oracle

In each pass, we have a candidate solution for LP9 and compute $y_{ij}$ based on the candidate. However we cannot store these $y_{ij}$ values. We use the following result:

**Theorem 25.** *[1] There exists a single pass semi-streaming algorithm that returns a sparsification of size $\tilde{O}(n/\epsilon^2)$ that estimates all cut values within $(1 \pm \epsilon)$ factor.*

We generate a stream of $y_{ij}$ as we stream through the edges, and construct a sparsification $\hat{y}_{ij}$ such that for any $U$, $(1 - \delta/C) \sum_{i \in U, j \notin U} y_{ij} \leq \sum_{i \in U, j \notin U} \hat{y}_{ij} \leq (1 + \delta/C) \sum_{i \in U, j \notin U} y_{ij}$ where $C$ is a constant to be defined later. We prove two lemmas which show that we can approximate LP10 given $\hat{y}_{ij}$ instead of $y_{ij}$.

**Lemma 26.** *Suppose that $(\mathbf{x}, \mathbf{z})$ is an optimal solution of LP10 given $y_{ij}$ and its objective value is $\beta$. Then, the objective value of $(\mathbf{x}, \mathbf{z})$ given $\hat{y}_{ij}$ is at least $(1 - \delta/2)\beta$.*

*Proof.* If $\sum_{i,j \in U} y_{ij} < \frac{1}{4} \sum_{i \in U} \sum_j y_{ij}$, $z_U = 0$. Suppose that it is not true. Then, we can increase $x_i$ by $\frac{1}{3} z_U$ for $i \in U$ and set $z_U = 0$. Then, left hand side of each constraints in LP10 decreases and therefore the modified solution is still feasible. On the other hand, the objective value increases by $z_U \left[ \frac{1}{3} \sum_{i \in U} \sum_j y_{ij} - \sum_{i,j \in U} y_{ij} \right] > 0$.

Now consider the case $\sum_{i,j \in U} y_{ij} \geq \frac{1}{4} \sum_{i \in U} \sum_j y_{ij}$, we have:

$$\sum_{i,j \in U} \hat{y}_{ij} = \frac{1}{2} \left[ \sum_{i \in U} \sum_j \hat{y}_{ij} - \sum_{i \in U} \sum_{j \notin U} \hat{y}_{ij} \right] \geq \frac{1}{2} \left[ \sum_{i \in U} \sum_j (1 - \delta/C) y_{ij} - (1 + \delta/C) \sum_{i \in U} \sum_{j \notin U} y_{ij} \right]$$

$$= \frac{1}{2} \left[ \sum_{i \in U} \sum_j y_{ij} - \sum_{i \in U} \sum_{j \notin U} y_{ij} \right] - \frac{\delta}{C} \sum_{i \in U} \sum_j y_{ij} \geq \sum_{i,j \in U} y_{ij} - \frac{4\delta}{C} \sum_{i,j \in U} y_{ij}.$$

For $C \geq 8$, we have $\sum_{i,j \in U} \hat{y}_{ij} \geq (1 - \delta/2) \sum_{i,j \in U} y_{ij}$ and $\sum_j \hat{y}_{ij} \geq (1 - \delta/2) \sum_j y_{ij}$ for all $i$. Therefore, the objective value of $(\mathbf{x}, \mathbf{z})$ given $\hat{y}_{ij}$ is at least $(1 - \delta/2)\beta$. □

**Lemma 27.** *Suppose that $(\mathbf{x}, \mathbf{z})$ is a solution of LP10 with the objective value less than $(1 - \delta)\beta$. Then, the objective value of $(\mathbf{x}, \mathbf{z})$ given $\hat{y}_{ij}$ is at most $(1 - \delta/2)\beta$.*

*Proof.* Suppose that the objective value of $(\mathbf{x}, \mathbf{z})$ given $y_{ij}$ is $(1 - q)\beta$. From the proof of Lemma 26, if $\sum_{i,j \in U} y_{ij} < \frac{1}{4} \sum_{i \in U} \sum_j y_{ij}$, we can increase the objective value by at least $z_U \frac{1}{12} \sum_{i \in U} \sum_j y_{ij}$ by modifying values of $\mathbf{x}, \mathbf{z}$. Therefore, $\sum_U z_U \frac{1}{12} \sum_{i \in U} \sum_j y_{ij} \leq q\beta$ for such sets $U$. From the sparsification,

$$\sum_{i,j \in U} \hat{y}_{ij} = \frac{1}{2} \left[ \sum_{i \in U} \sum_j \hat{y}_{ij} - \sum_{i \in U} \sum_{j \notin U} \hat{y}_{ij} \right] \leq \frac{1}{2} \left[ (1 + \delta/C) \sum_{i \in U} \sum_j y_{ij} - (1 - \delta/C) \sum_{i \in U} \sum_{j \notin U} y_{ij} \right]$$

$$\leq \frac{1}{2} \left[ \sum_{i \in U} \sum_j y_{ij} - \sum_{i \in U} \sum_{j \notin U} y_{ij} \right] + \frac{\delta}{C} \sum_{i \in U} \sum_j y_{ij} = \sum_{i,j \in U} y_{ij} + \frac{\delta}{C} \sum_{i \in U} \sum_j y_{ij}.$$

So the error term derived by such $U$ is at most $\frac{12\delta q}{C}\beta$. If $\sum_{i,j\in U} y_{ij} \geq \frac{1}{4}\sum_{i\in U}\sum_j y_{ij}$, $\sum_{i,j\in U} \hat{y}_{ij} \leq (1+4\delta/C)\sum_{i,j\in U} y_{ij}$. We also have $\sum_j \hat{y}_{ij} \leq (1+\delta/C)\sum_j y_{ij}$. From these three inequalities, the objective value of $(\mathbf{x},\mathbf{z})$ given $\hat{y}_{ij}$ is at most $(1+4\delta/C)(1-q)\beta + (12\delta q/C)\beta$. For a sufficiently large $C$ and $q > \delta$, the value is at most $(1-\delta/2)\beta$. □

## 4.3 Solving the Oracle

In this section we discuss how to solve the oracle described in LP10. We will be using the multiplicative weights update method (see Section 2.4). This method solves the dual problem explicitly and produces a certificate of optimality for the primal. Therefore we will apply the method to the dual of LP10, and have an explicit solution of the dual of the dual of LP10 (which is LP10). The dual of LP10 is described in LP11.

$$\begin{aligned}
\min \quad & \sum_i 3\zeta_i + \zeta_\alpha \\
\text{s.t} \quad & 2\zeta_i + \tfrac{1}{\alpha}\zeta_\alpha \geq \sum_j y_{ij} && \forall i \quad \text{(Type-A)} \\
& \sum_{i\in U} \zeta_i + \lfloor \tfrac{|U|}{2} \rfloor \tfrac{1}{\alpha}\zeta_\alpha \geq \sum_{i,j\in U} y_{ij} && \forall U \quad \text{(Type-B)}
\end{aligned}$$

If the number of nodes in the graph is even, we first add a node with no edges adjacent to it and use the multiplicative weights update method. We will explain the reason why we add the node later. In each iteration of the multiplicative weights update method, we are given weights $\zeta_i$ for each node $i$ and $\zeta_\alpha$ which correspond to the constraints of LP10. We find a violated constraint of LP11 and construct a dual witness with it. Algorithm 8 is the oracle for LP11.

---

**Algorithm 8** Oracle for LP11.

1: Normalize $\zeta_i$ and $\zeta_\alpha$ so that $\sum_i 3\zeta_i + \zeta_\alpha = \beta$.
2: **if** there exists a node $i$ such that $2\zeta_i + \tfrac{1}{\alpha}\zeta_\alpha < \sum_j y_{ij}$ and $\sum_j y_{ij} > \delta\beta/\alpha$ **then**
3:     Return $x_i = \beta/\sum_j y_{ij}$ (and 0 for other variables).
4: **end if**
5: $\zeta_\alpha \leftarrow \zeta_\alpha + \delta\beta$. (Ensures that all type-A constraints are satisfied; since we can have $\sum_j y_{ij} \leq \delta\beta/\alpha$ above.)
6: $\zeta_\alpha \leftarrow (1+2\delta)\zeta_\alpha$. (Ensures that $U$ with $|U| > \tfrac{1}{\delta}$ is not used in line 11.)
7: Add a node $s$ to the graph where an edge $(i,s)$ has weight $\tfrac{1}{2}(2\zeta_i - \tfrac{1}{\alpha}\zeta_\alpha - \sum_j y_{ij})$.
8: Find a minimum odd cut $(S,U)$ with $s \in S$. Note that both $S$ and $U$ are odd sets.
9: **if** the minimum cut value is less than $(\zeta_\alpha - \delta\beta)/\alpha$ **then**
10:     Let $\Delta = \sum_{i,j\in U} y_{ij}$.
11:     Return $z_U = \beta/\Delta$ (and 0 for other variables).
12: **else**
13:     The oracle reports failure. We indicate the verification step: Increase $\zeta_\alpha$ by $\delta\beta$ and return $(\{\zeta_i\}, \zeta_\alpha)$ as the overall primal solution.
14: **end if**

---

**Lemma 28.** *Algorithm 8 returns an admissible solution with $\ell = 3$ and $\rho = O(n/\delta)$.*

*Proof.* It is obvious that $\mathbf{c}^T\mathbf{y} = \beta$ from line 3 and line 11. For the other conditions of admissibility, we show that we only pick a constraint where the right hand side is at least $\frac{\delta\beta}{\alpha}$. It is obvious in the case of type-A constraints. If there is no violated type-A constraint whose RHS is at least $\delta\beta/\alpha$, we satisfy all type-A constraints in line 5 by increasing $\zeta_\alpha$.

For an odd set $U$, we have

$$\sum_{i \in U} \zeta_i + \left\lfloor \frac{|U|}{2} \right\rfloor \frac{1}{\alpha} \zeta_\alpha - \sum_{i,j \in U} y_{ij} = \frac{1}{2} \sum_{i \in U} \left( 2\zeta_i + \frac{1}{\alpha} \zeta_\alpha \right) - \frac{1}{\alpha} \zeta_\alpha - \sum_{i,j \in U} y_{ij}$$

$$= \frac{1}{2} \sum_{i \in U} \left( 2\zeta_i + \frac{1}{\alpha} \zeta_\alpha - \sum_j y_{ij} \right) + \sum_{i \in U} \sum_{j \notin U} y_{ij} - \frac{1}{\alpha} \zeta_\alpha.$$

The first term is the weight of an edge between $s$ and $i \in U$ and the second term is the cut value of $U$ excluding edges to $s$. So if the cut value of $U$ is less than $\frac{1}{\alpha} \zeta_\alpha - \frac{\delta \beta}{\alpha}$, the constraint corresponding to $U$ is violated by at least $\frac{\delta \beta}{\alpha}$. Since the left hand side of a constraint is always positive, this proves that the right hand side is at least $\frac{\delta \beta}{\alpha}$.

In addition, if $|U| > \frac{1}{\delta}$, the corresponding cut value is greater than $(\zeta_\alpha - \delta \beta)/\alpha$. Let $\zeta'_\alpha$ and $\zeta_\alpha$ be the value of $\zeta_\alpha$ before and after line 6 since for $|U| > \frac{1}{\delta}$,

$$\sum_{i \in U} \left( 2\zeta_i + \frac{1}{\alpha} \zeta_\alpha - \sum_j y_{ij} \right) + \sum_{i \in U} \sum_{j \notin U} y_{ij} - \frac{1}{\alpha} \zeta_\alpha = \sum_{i \in U} \left( 2\zeta_i + \frac{1 + 2\delta}{\alpha} \zeta'_\alpha - \sum_j y_{ij} \right) + \sum_{i \in U} \sum_{j \notin U} y_{ij} - \frac{1 + 2\delta}{\alpha} \zeta'_\alpha$$

$$= \sum_{i \in U} \left( 2\zeta_i + \frac{1}{\alpha} \zeta'_\alpha - \sum_j y_{ij} \right)$$

$$+ \sum_{i \in U} \sum_{j \notin U} y_{ij} + \frac{2\delta |U|}{\alpha} \zeta'_\alpha - \frac{1 + 2\delta}{\alpha} \zeta'_\alpha$$

$$\geq 0.$$

So we assign at most $\frac{\alpha}{\delta}$ to $x_i$ or $z_U$ (with $|U| \leq \frac{1}{\delta}$) and the values are all positive. Also only one variable has positive value. For $\zeta_i$, we have $\mathrm{M}(i, \mathbf{y}^t) = 2x_i + \sum_{i \in U} z_U - 3$ and for $\zeta_\alpha$, we have $\mathrm{M}(\alpha, \mathbf{y}^t) = \frac{1}{\alpha} \left[ \sum_i x_i + \sum_U \lfloor \frac{|U|}{2} \rfloor z_U \right] - 1$. Then, $-3 \leq \mathrm{M}(i, \mathbf{y}^t) \leq 2\alpha/\delta$ and $-1 \leq \mathrm{M}(\alpha, \mathbf{y}^t) \leq \delta n/2$

Since we only choose a violated constraint, we have the following inequality.

$$\mathrm{M}(\mathcal{D}^t, \mathbf{y}^t) = \frac{1}{\zeta_\alpha + \sum_i \zeta_i} \left[ \zeta_\alpha \mathrm{M}(\alpha, \mathbf{y}^t) + \sum_i \zeta_i \mathrm{M}(i, \mathbf{y}^t) \right]$$

$$= \frac{1}{\zeta_\alpha + \sum_i \zeta_i} \left[ \zeta_\alpha \left[ \sum_i x_i + \sum_U \left\lfloor \frac{|U|}{2} \right\rfloor - 1 \right] + \zeta_i \left( \sum_i 2x_i + \sum_{i \in U} z_U - 3 \right) \right]$$

$$= \frac{1}{\zeta_\alpha + \sum_i \zeta_i} \left[ \sum_i x_i \left( 2\zeta_i + \frac{1}{\alpha} \zeta_\alpha \right) + \sum_U z_U \left( \sum_{i \in U} \zeta_i + \left\lfloor \frac{|U|}{2} \right\rfloor \frac{1}{\alpha} \zeta_\alpha \right) - \left( \sum_i 3\zeta_i + \zeta_\alpha \right) \right]$$

$$< \frac{1}{\zeta_\alpha + \sum_i \zeta_i} \left[ \sum_i x_i \left( \sum_j y_{ij} \right) + \sum_U z_U \left( \sum_{i,j} y_{ij} \right) - \beta \right]$$

$$= \frac{1}{\zeta_\alpha + \sum_i \zeta_i} \left[ \sum_{i,j} y_{ij} \left( x_i + x_j + \sum_{i,j \in U} z_U \right) - \beta \right] = O(\delta).$$

□

**Lemma 29.** *If the oracle reports failure, it returns a feasible solution for LP11 with value at most $(1 + 5\delta)\beta$ is returned.*

*Proof.* If a type-A constraint is violated more than $\delta\beta/\alpha$, the oracle returns a dual witness. Other constraints are satisfied by line 5 in which the objective value is increased by $\delta\beta$. In line 6, we increase the objective value by at most $(1+2\delta)$ factor. After that, if a type-B constraint is violated more than $\delta\beta/\alpha$, the oracle returns a dual witness. If not, the constraints are satisfied by line 13 and the objective value is increased by $\delta\beta$. Initially the objective value is $\beta$. So the oracle returns a feasible solution for LP11 with value at most $(1+2\delta)(1+\delta)\beta + \delta\beta \leq (1+5\delta)\beta$ (for sufficiently small $\delta$). □

## 4.4 Storing $z_U$ and Estimating $\sum_{i,j \in U} z_U$ (to compute $y_{ij}$)

From Lemma 28, Lemma 29, and Theorem 4, we can find a $(1-\delta)$-approximation of LP10 in $\tilde{O}(\frac{n^2}{\delta^2})$ iterations. However, the size of the output for LP10 can be superlinear in the number of nodes.

Note that the analysis in Section 4.3 requires the estimation of $\sum_{i,j \in U} z_U$ with $\delta$ additive error not the exact value of $z_U$ for all $U$. In this section, we present an algorithm that produces a sketch of $z_U$ that approximates $\sum_{i,j \in U} z_U$ within $\delta$ additive error for all $i, j$ pairs. We use the Johnson-Lindenstrauss lemma [15]. Especially we use the fact that we can construct a random linear mapping for the Johnson-Lindenstrauss lemma obliviously [6].

**Theorem 30.** *[6] Suppose that $0 < \epsilon < 1$, $n$ is a integer, and $k = \lceil \frac{C}{\epsilon^2} \log n \rceil$ with a constant $C$. Let $V$ be a set of $n$ points in $\mathbb{R}^d$ and let $A$ be a $d \times k$ matrix such that $A_{ij} \sim \frac{1}{\sqrt{k}} N(0, 1)$. Then, $(1 - \epsilon)\|u - v\|^2 \leq \|Au - Av\|^2 \leq (1 + \epsilon)\|u - v\|^2$ for all $u, v \in V$ with high probability.*

Note that the construction of $A$ is oblivious to $V$, i.e., we can construct $A$ without reading $V$. Let $U_1, U_2, \cdots, U_t$ be the sets we chose at time $t$ and $z_t$ be the weight we give to $U_t$. Let $X_i$ be a vector such that
$$X_{it} = \begin{cases} \sqrt{z_t} & \text{if } i \in U_t \\ 0 & \text{otherwise} \end{cases}$$

Then $\sum_{i,j \in U} z_U = (\|X_i\|^2 + \|X_j\|^2 - \|X_i - X_j\|^2)/2$. For each $i$, we store $AX_i$ instead of $X_i$ and $A$ can be generated column by column as we proceed. Then, we estimate $\|X_i\|^2$ and $\|X_i - X_j\|^2$ within $1 \pm O(\delta)$ factor. Since $\|X_i\|^2 = \sum_{i \in U} z_U \leq 3$, the multiplicative error translates into $O(\delta)$ additive error in $\|X_i\|^2 + \|X_j\|^2 - \|X_i - X_j\|^2$. The total space required is $O(\frac{n}{\delta^2} \log n)$.

## 4.5 Proof of Theorem 22

*Proof.* The sparsification step requires $\tilde{O}(\frac{n}{\delta^2})$ space (and $\tilde{O}(m)$ time). The space required for the sketch of $z_U$ is also $\tilde{O}(\frac{n}{\delta^2})$ since we do not have to remember the random matrix used for the Johnson-Lindenstrauss lemma. The space requirement of this step is dominated by the space requirement for the sampled edges.

The sampling and the construction of the sparsification can be done in $\tilde{O}(\frac{m}{\delta^2})$ time. We solve LP10 in $\tilde{O}(\frac{n}{\delta^3})$ iterations. The minimum odd cut algorithm takes $\tilde{O}(n^2 m')$ time as mentioned in the proof of Theorem 22 where $m'$ is the number of edges. The number of edges in the sparsification cab be reduced to $\tilde{O}(\frac{n}{\delta^2})$ by applying the sparsification algorithm again. So finding a minimum odd cut takes $\tau = \tilde{O}(\frac{n^3}{\delta^2})$ time. Note that the time to construct sparsifications is dominated by the minimum odd cut algorithm. We need $O(\frac{1}{\delta^2} \log n)$ iterations of the violation oracle. So the algorithm takes $\tilde{O}(\frac{n\tau}{\epsilon^5})$ time in total (for a particular guess of optimum solution $\alpha$). Since we execute the algorithm for $O(\frac{1}{\epsilon})$ guesses of the optimal solution, we have additional $\frac{1}{\epsilon}$ term in the space and time. □

## 4.6 Space-Pass Tradeoffs and Theorem 23

The sparsification algorithm in [1] is based on sampling of edges. The sampling probability of an edge depends on the edge's weight and strong-connectivity where the strong-connectivity is defined as follows.

**Definition 9.** *[4] A node-induced subgraph $G'$ is $k$-strong connected if the size of a minimum cut in $G'$ is $k$. An edge $e$ is $k$-strong connected if $e$ is in a $k$-strong connected component.*

If an edge is $k$-strong connected and has weight $y$, its sampling probability is $\Omega(\frac{y \log n}{\delta^2 k})$. The proofs in [1, 4] hold even if we increase the sampling probability. Hence, if we know the upperbound of weight and the lowerbound of strong connectivity for every edge, we can process the sampling step first and construct the actual sparsification after the edge weights are fixed. This can be used for processing multiple iterations in one pass.

**Lemma 31.** *In Algorithm 1, let $\mathbf{x}$ be the primal candidate in time $t$ and $\mathbf{x}'$ be the primal candidate after the update, i.e., $\mathbf{x}' = (1-\sigma)\mathbf{x} + \sigma\tilde{\mathbf{x}}$. Let $\mathbf{y}$ and $\mathbf{y}'$ be the corresponding dual candidates. For any constraint,*
$$\exp(-\delta/2) \leq \frac{\mathbf{y}_i}{\mathbf{y}'_i} \leq \exp(\delta/2)$$

*Proof.* Note that, $\frac{\mathbf{y}_i}{\mathbf{y}'_i} = \exp(\kappa \mathbf{A}_i(\sigma\mathbf{x} - \sigma\tilde{\mathbf{x}})/\mathbf{b}_i) = \exp(\kappa\sigma \mathbf{A}_i(\mathbf{x} - \tilde{\mathbf{x}})/\mathbf{b}_i)$. By the definition of width parameter, $-2\rho \leq \mathbf{A}_i(\mathbf{x} - \tilde{\mathbf{x}})/\mathbf{b}_i \leq 2\rho$. Since $\sigma = \frac{\delta}{4\kappa\rho}$, we obtain the desired result. □

Therefore, the edge weight (and the strong connectivity) increases or decreases by at most $\exp(k\delta/2)$ factor in $k$ iterations which leads to at most $\exp(k\delta)$ factor increase in the sampling probability. For each pass, we sample $k$ sets of edges which we will use to construct sparsifications for $k$ iterations. The sampling probability is $\exp(k\delta)$ times the sampling probability in [1]. Once we have such sampled edges, we can process $k$ iterations without reading the data stream. Now we prove Theorem 22.

*Proof.* (Of Theorem 23.) Since we increase the sampling probability by $\exp(k\delta)$, in the sparsification, it increases the size of sampled edge set by $\exp(k\delta)$ factor. Since we have $k$ such sets, the space requirement for the sparsification is $\tilde{O}(k \exp(k\delta) n/\delta^2)$ and the number of passes is $O(\frac{\log n}{\delta^2 k})$. By using $k = \frac{\log n}{p\delta}$ and $\delta = \Theta(\epsilon)$, and following the rest of the proof of Theorem 22, we obtain the space and the number of passes. □

## 5 Estimating the Size of the Maximum Weighted Matching

In this section, we estimate the size of the maximum weighted matching in general graphs by applying the algorithm in Section 4 to the maximum weighted matching. The dual linear program formulation for the maximum weighted matching is as follows:

$$\begin{aligned}
\min \quad & \sum_i x_i + \sum_U \lfloor \tfrac{|U|}{2} \rfloor z_U \\
\text{s.t} \quad & x_i + x_j + \sum_{i,j \in U} z_U \geq w_{ij} \quad \forall (i,j) \in E \\
& x_i \geq 0
\end{aligned} \qquad \text{(LP12)}$$

Edge weights introduce complications which prevent direct application of the fractional covering formulation and the corresponding oracle of Section 4. In particular the proof of Corollary 24 does not apply to weighted graphs. We proceed in the following three steps.

1. We first run a filtering step to eliminate very small weights. This reduces the width of the inner step, or the violation oracle, such that we can solve the oracle in small space and small number of passes.

2. We then run a second filtering step to ensure that the ratio of the largest to smallest weights incident to a vertex are bounded. This reduces the width of the outer step, or the overall fractional covering framework, such that the number of passes are bounded.

3. We then slightly modify the LP formulation for the unweighted case in Section 4.

As a consequence we achieve the following result:

**Theorem 32.** *We can approximate the size of the maximum weighted matching in $\tilde{O}(\frac{p}{\epsilon^6})$ passes, $\tilde{O}(\frac{n^{1+1/p}}{\epsilon^4})$ space, and $\tilde{O}(\frac{n^4}{\epsilon^{14}})$ time.*

We present the three steps and the proof of Theorem 32 in the remainder of the section. We first show how to reduce the inner width (step 1). We then discuss the formulation we need to solve (step 3), and based on the formulation we show how to achieve the second preprocessing (in step 2) to reduce the width of the fractional covering framework.

## 5.1 Preprocessing I: Filtering Small Weight Edges

The width parameter of the violation oracle is $O(\alpha/\delta)$ which is $O(n/\delta)$ for MCM. However, $\alpha$ could be arbitrarily large (compared to the minimum edge weight) for MWM. We modify the input graph so that $\alpha = O(n/\delta)$ and the minimum edge weight is 1.

We first execute an one-pass $O(1)$-approximation algorithm on the graph and compute the lowerbound of the optimal solution. Let this value be $A$. We delete any edge whose weight is less than $\delta A/n$. The weight of the optimal solution decreases by at most $\delta A$. Since the $\alpha = O(A)$ and the minimum edge weight is $\delta A/n$, by scaling the edge weights, we obtain the desired graph. The width parameter of the violation oracle would be $O(n/\delta^2)$ compared to $O(n/\delta)$ in Section 4.3

## 5.2 The Modified Fractional Covering Formulation

Let $\Phi_i = \max_j w_{ij}$ and $\Psi_i = \min_j w_{ij}$. We use the following fractional covering formulation.

$$\begin{array}{ll} \max & \lambda \\ \text{s.t} & x_i + x_j + \sum_{i,j \in U} z_U \geq \lambda w_{ij} \quad \forall (i,j) \in E \\ P: & \sum_i x_i + \sum_U \lfloor \frac{|U|}{2} \rfloor z_U \leq \alpha \\ & 2x_i + \sum_{i \in U} z_U \leq \frac{2}{\delta} \Phi_i \quad \forall i \end{array} \quad \text{(LP13)}$$

We first show that the above formulation is valid, i.e., it approximates the optimal solution of LP12.

**Lemma 33.** *Let $OPT$ be the optimal solution for LP12. Then, there exists a feasible solution (with $\lambda = 1$) for LP13 whose objective value is at most $(1+\delta)OPT$.*

*Proof.* We start with an optimal solution $(\mathbf{x}, \mathbf{z})$ of LP12 and construct a feasible solution for LP13. Let $i$ be a node such that $2x_i + \sum_{i \in U} z_U > \frac{2}{\delta} \Phi_i$. For each $U$ with $i \in U$, we increase $x_j$ with $j \in U, j \neq i$ by $\frac{z_U}{2}$ and set $z_U$ to be zero. For $i$, we assign $\Phi_i$ to $x_i$. For edges adjacent to $i$, the edge is satisfied by the value of $x_i$. For edges $(i', j)$ not adjacent to $i$, we increase $x_{i'} + x_j$ by $z_U$ while set $z_U$ to be zero. So the constraint is still satisfied after the change.

The process increase the objective value by $\Phi_i$. But it decreases $\sum_U z_U$ by at least $\frac{2}{\delta}\Phi_i$. Since $\sum_U z_U$ is less than $OPT$ initially, the total increase in the objective value is bounded by $\delta OPT$. Therefore, there exists a feasible solution for LP13 whose objective value is at most $(1+\delta)OPT$. □

## 5.3 Preprocessing II: Reducing the Width of LP13

The width parameter of LP13 is $\frac{1}{\delta}\max_i \frac{\Phi_i}{\Psi_i}$. The parameter depends on the largest weight ratio between two edges that share an endpoint. Algorithm 9 finds a subgraph $G'$ that contains a sufficiently large matching while the width parameter is bounded. The algorithm is a modification of Algorithm 7 in [2].

---
**Algorithm 9** Finding a subgraph $G'$
---
1: **for each tier** $k = 0, 1, \cdots, L$ **in parallel do**
2:     Find a maximal matching.
3:     Let $C_k$ be the set of nodes matched in the maximal matching.
4:     Let $S_k^1 = C_k$
5:     **for** $t = 1$ **to** $q$ **do**
6:        Find a maximal matching between $C_k$ and $V - S_k^t$.
7:        Let $T_k^t$ be the set of nodes matched in the maximal matching.
8:        $S_k^{t+1} = S_k^t \cup T_k^t$.
9:     **end for**
10: **end for**
11: Let $u_i^1 = (1+\delta)^k$ for the maximum $k$ with $i \in S_k^q$.
12: Return $G' = (V, E')$ with $E' = \{(i,j): \frac{\delta^2}{q}u_i, \frac{\delta^2}{q}u_j \leq w_{ij} \leq u_i, u_j\}$.

---

**Lemma 34.** *The width parameter of LP13 is bounded by $\frac{q}{\delta^3}$.*

*Proof.* Follows from inspection of line line 12. □

We now show that $G'$ has a sufficiently large matching for a suitable setting of $q$.

**Lemma 35.** *Let $OPT$ be the size of maximum weighted matching and $q = O(\frac{1}{\delta^2} \log \frac{1}{\delta})$. $G'$ contains a matching of size $(1 - O(\delta))OPT$.*

*Proof.* First, we eliminate all edges $(i,j)$ such that $w_{ij} > u_i$ or $w_{ij} > u_j$ and let $G''$ be the resulting subgraph of $G$. $G''$ contains a matching of weight at least $(1 - O(\delta))OPT$ [2]. Now we fix an optimal solution in $G''$ and eliminate all edges $(i,j)$ such that $w_{ij} < \frac{\delta^2}{q}u_i$ or $w_{ij} < \frac{\delta^2}{q}u_j$. For each $i \in C_k$, we pick at most $q$ neighbors in tier $k$ which eliminates at most $\delta^2(1+\delta)^k$ weight. Let $(i,j)$ be the edge matched in the initial maximal matching of tier $k$. Then, at least one of $i$ and $j$ must be matched to an edge with weight at least $(1+\delta)^k/2$ in the optimal solution since otherwise we can improve the optimal solution by replacing the edge. Let this edge be $e$. We charge the weight of edges that are eliminated by $i$ to $e$. Each edge $e$ in the optimal solution can be charged at most

$$2\sum_{i=0}^{\infty} \frac{\delta^2 w_e}{(1+\delta)^i} \leq 2\delta w_e.$$

Therefore, we obtain a matching of size at least $(1 - O(\delta))OPT$. □

## 5.4 Proof of Theorem 32

The approximation ratio is guaranteed by the discussion in Section 5.1 and Lemmas 33 and 35.

The width parameter is $\frac{1}{\delta^5} \log \frac{1}{\delta}$ based on Lemma 34. The rest of the algorithm and analysis is identical to Section 4 (except that the multiplicative weight update step has width $n/\delta^2$ instead of $n/\delta$). Thus we obtain the Theorem 32.